\documentclass[fleqn,usenatbib]{mnras}
\usepackage[T1]{fontenc}
\usepackage[latin9]{inputenc}
\usepackage{newtxtext}
\usepackage{orcidlink}
\usepackage{color}
\usepackage{babel}
\usepackage{amstext}
\usepackage{amssymb}
\usepackage{amsmath}
\usepackage{graphicx}

\newcommand{\lyxdot}{.}
\usepackage{color}
\usepackage{babel}

\title{Influence of turbulence on Lyman-alpha scattering}

\author[V. R. Munirov and A. A. Kaurov]{
Vadim R. Munirov\orcidlink{0000-0001-6711-1272},$^{1}$\thanks{E-mail: vmunirov@berkeley.edu}
Alexander A. Kaurov\orcidlink{0000-0003-0255-1204}$^{2,3,4}$
\\
$^{1}$Department of Physics, University of California, Berkeley, CA
94720, USA\\
$^{2}$Department of the History of Science, Harvard University, Cambridge,
MA 02138, USA\\
$^{3}$Program in Interdisciplinary Studies, Institute for Advanced Study,
Princeton, NJ 08540, USA\\
$^{4}$Blue Marble Space Institute of Science, Seattle, WA 98104, USA
}

\date{Accepted 2023 April 14. Received 2023 March 29; in original form 2022 August 30}

\pubyear{2023}

\begin{document}
\label{firstpage}
\pagerange{\pageref{firstpage}--\pageref{lastpage}}
\maketitle

\begin{abstract}
We develop a Monte Carlo radiative transfer code to study the effect
of turbulence with a finite correlation length on scattering of Lyman-alpha
(Ly$\alpha$) photons propagating through neutral atomic hydrogen
gas. We investigate how the effective mean free path, the emergent
spectrum, and the average number of scatterings that Ly$\alpha$ photons
experience change in the presence of turbulence. We find that the
correlation length is an important and sensitive parameter that has an influence on physically relevant properties of Ly$\alpha$ radiative transfer. In particular, it can significantly, by orders of magnitude, reduce the number of scattering
events that the average Ly$\alpha$ photon undergoes before it escapes
the turbulent cloud.
\end{abstract}

\begin{keywords}
radiative transfer -- scattering -- turbulence -- galaxies: high-redshift
\end{keywords}

\section{Introduction}

The resonant Lyman-alpha (Ly$\alpha$) scattering generates a number
of phenomena that are of interest in astrophysics and cosmology. Ly$\alpha$
emitters are a unique and powerful tool to learn about structure formation
during the epoch of reionization \citep{McQuinn2007,Dijkstra2014,Hayes2015,Behrens2019,Ouchi2020}.
Studying Ly$\alpha$ emission, propagation, and absorption allows
one to constrain the ionization state of the universe, probe galaxies
kinematics and dynamics, and deduce properties of the atomic gas surrounding
them \citep{Erb2018,Hayes2015,Wolfe2005,Ouchi2020}.

The resonant scattering of Ly$\alpha$ photons through regions of
neutral hydrogen ($\textrm{HI}$ regions) of the intergalactic (IGM)
and interstellar (ISM) media is a process which is random both in
frequency and physical spaces. Except for certain cases when analytical
solutions are available \citep{Adams1972,Harrington1973,Neufeld1990,Loeb1999,Dijkstra2006a},
one has to resort to numerical methods to study the propagation of
Ly$\alpha$ radiation in these systems. The most efficient and hence
popular way is to use Monte Carlo radiative transfer simulations \citep{Dijkstra2006a,Dijkstra2006b,Dijkstra2014,Laursen2009,Laursen2010,Zheng2002,Zheng2014,Semelin2007,Smith2018,Smith2020,Seon2020,Seon2022}.
The influence of many effects on the propagation of Ly$\alpha$ photons
has been studied, but the effect of unresolved turbulence has received
much less attention. Meanwhile, at least in some of the Ly$\alpha$
emitters the photon escape is believed to be driven by turbulence
in the star-forming gas \citep{Puschnig2020_XI}. The extraction of the ISM turbulence parameters from the observations requires combination of different techniques --  statistical methods and comparison with the simulations -- that allow for disentanglement of the velocity and density information. The most common statistical parameters used for describing the turbulence are the slopes of velocity and density power spectra. These numbers vary from -2.0 to -1.4 and from -1.6 to -0.4 correspondingly for the observed systems (e.g. \citealt{Burkhart2021}). The power law behavior cuts off at Kolmogorov length scale, for instance in the observation of the Taurus molecular cloud the HI velocity spectrum cuts off at 0.3 pc, but the turbulent cascade continues in H2 \citep{Yuen2022}.

Usually turbulence in Monte Carlo radiative transfer codes is treated
as an effective thermal velocity that changes the Doppler broadening
parameter from $b=v_{\textrm{th}}$ to $b=\sqrt{v_{\textrm{th}}^{2}+v_{\textrm{turb}}^{2}}$~\citep{Eastman1985,Jaffel1993,Smith2021}.
This corresponds to the so-called model of microturbulence. However,
besides its amplitude ($v_{\textrm{turb}}$), turbulence has another
crucial parameter \textendash{} the correlation length ($l_{\textrm{turb}}$). Such turbulence with a finite correlation length is often referred to as macroturbulence. In principle, it is possible to execute Monte Carlo radiative
transfer codes along with hydrodynamic simulations \citep{Smith2020} which provide the macroscopic velocity field describing gas motion up until a certain resolution scale. Then, the turbulence below the resolution scale is usually treated as microturbulence. However, such an approach can be too numerically costly. Moreover, it is not clear what resolution scale should be chosen to justify the use of microturbulence model below the resolution scale. Thus,
there is a need in simplified description of turbulent macroscopic
motion.
It is the goal of this paper to provide this simplified model of macroturbulence and investigate the influence of a finite
turbulence correlation length on Ly$\alpha$ scattering. To accomplish
this we have written an efficient Monte Carlo radiative transfer code
using \texttt{Python 3} \citep{Python3} sped up with \texttt{Cython}
\citep{Cycthon} and \texttt{Numba} \citep{Numba} that accounts for
a simple model of turbulence with a finite correlation length. We
find that the correlation length is an important parameter that affects
the emergent spectrum and can change the number of scatterings that
the average photon experiences before it escapes by orders of magnitude.

The paper is organized as follows. In Section~\ref{Section_2}, we
describe the Monte Carlo model of the propagation of Ly$\alpha$ radiation
through turbulent gas that we developed. In Section~\ref{Section_3},
we present the results of numerical simulations for the emergent spectrum
and the number of scatterings for Ly$\alpha$ photons traveling though
a turbulent cloud of neutral hydrogen gas. Finally, in Section~\ref{Section_4},
we summarize and discuss the possible implications of the obtained
results.

\begin{figure*}
\includegraphics[width=0.5\textwidth]{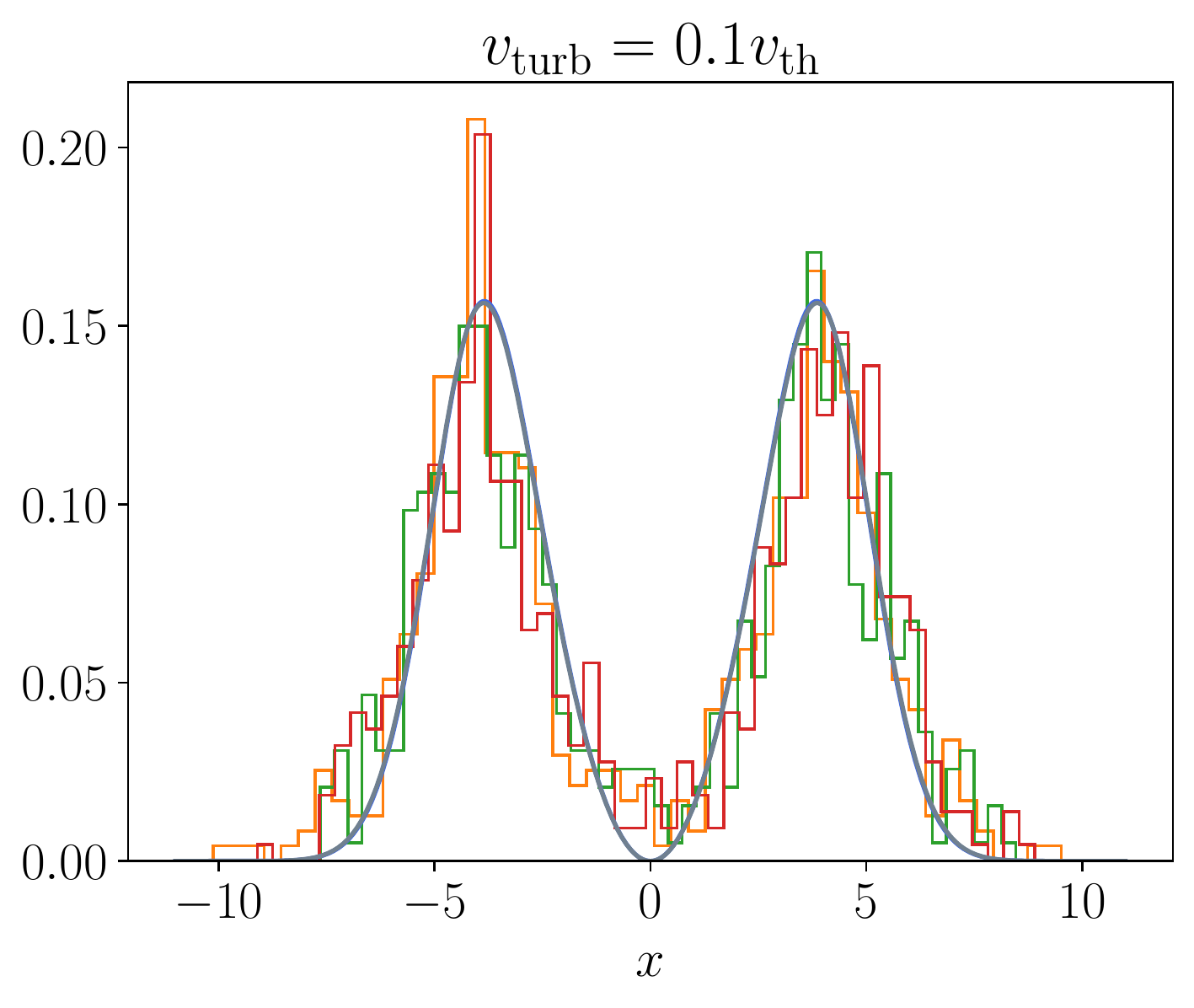}\includegraphics[width=0.5\textwidth]{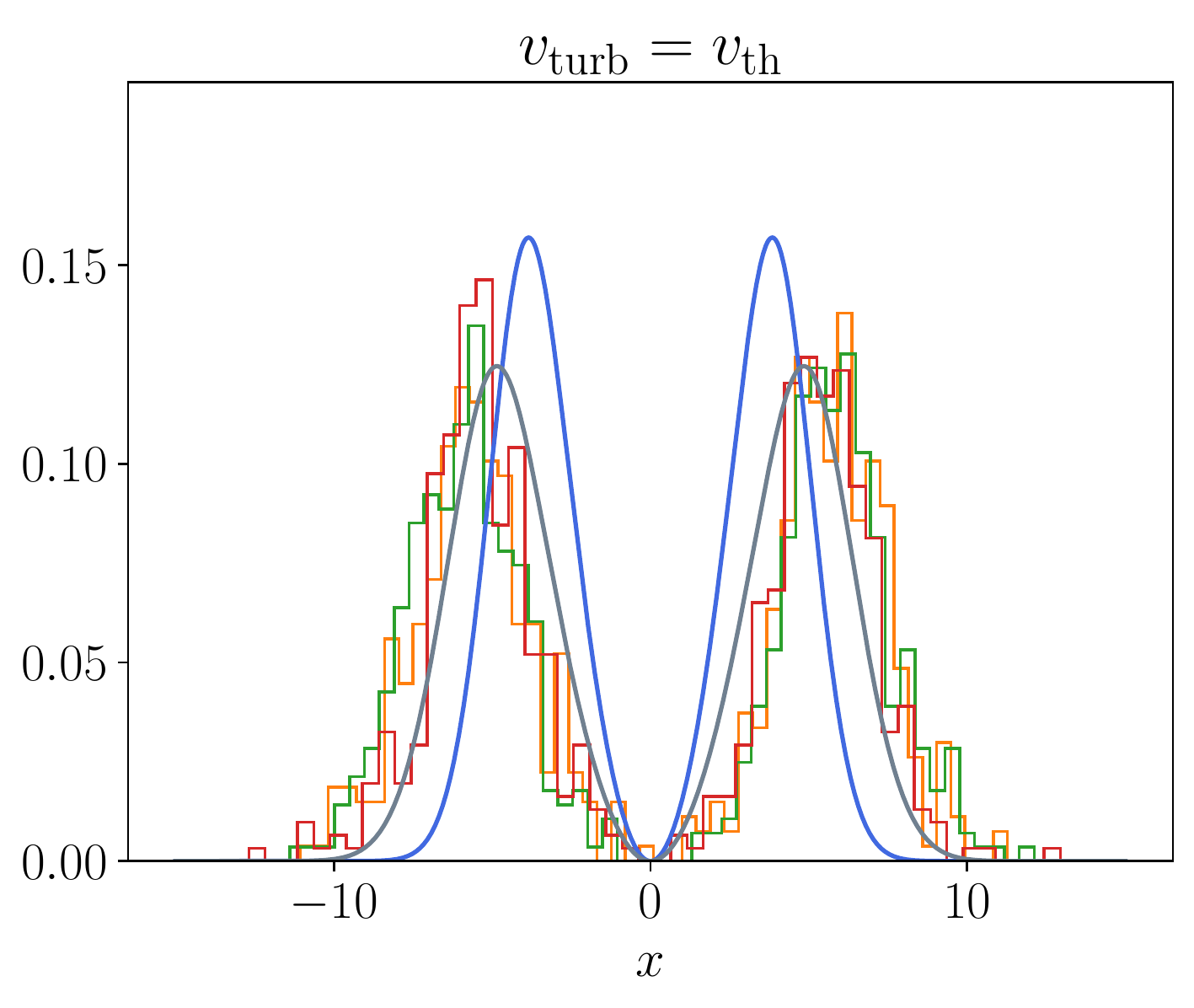}

\includegraphics[width=0.5\textwidth]{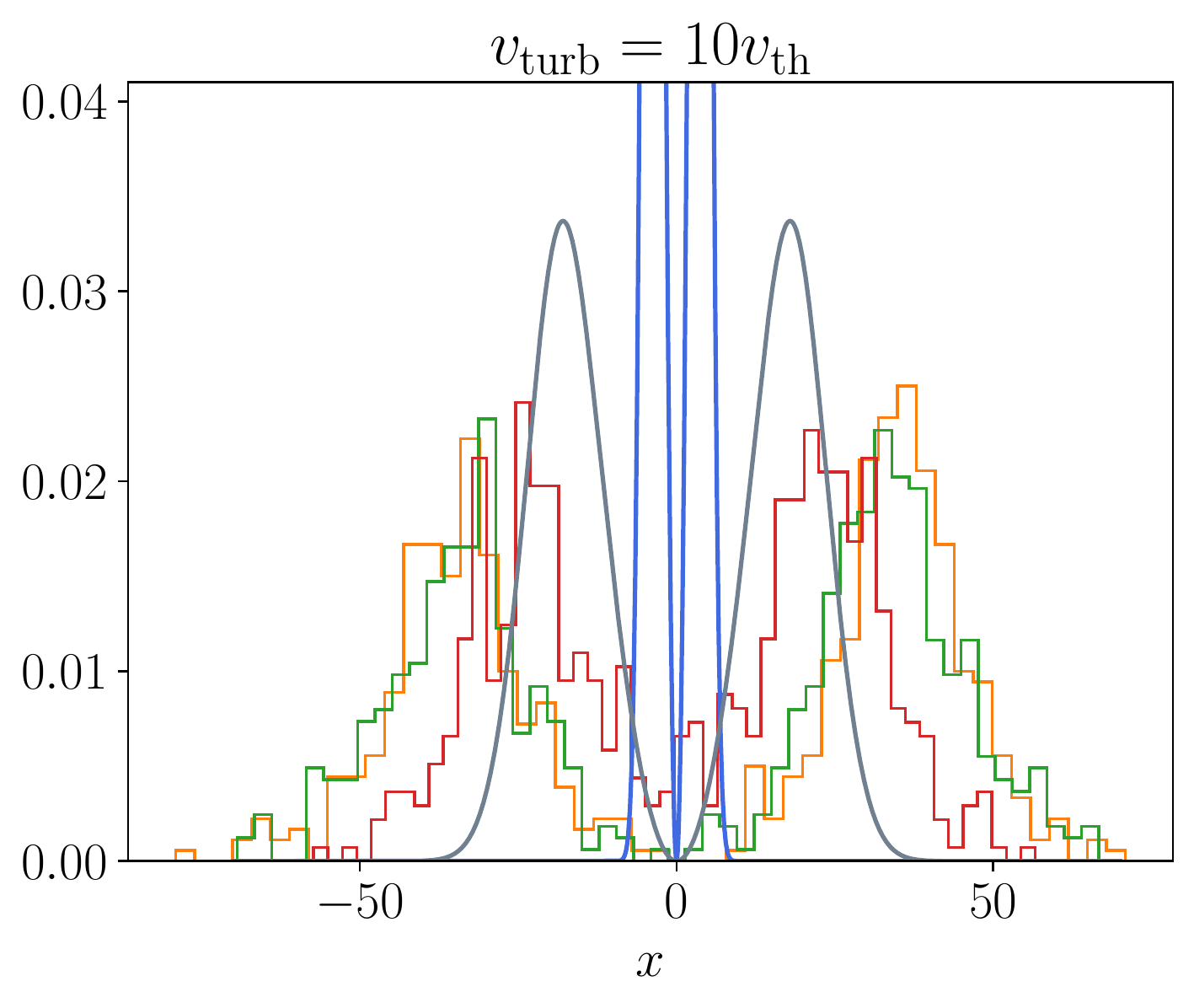}\includegraphics[width=0.5\textwidth]{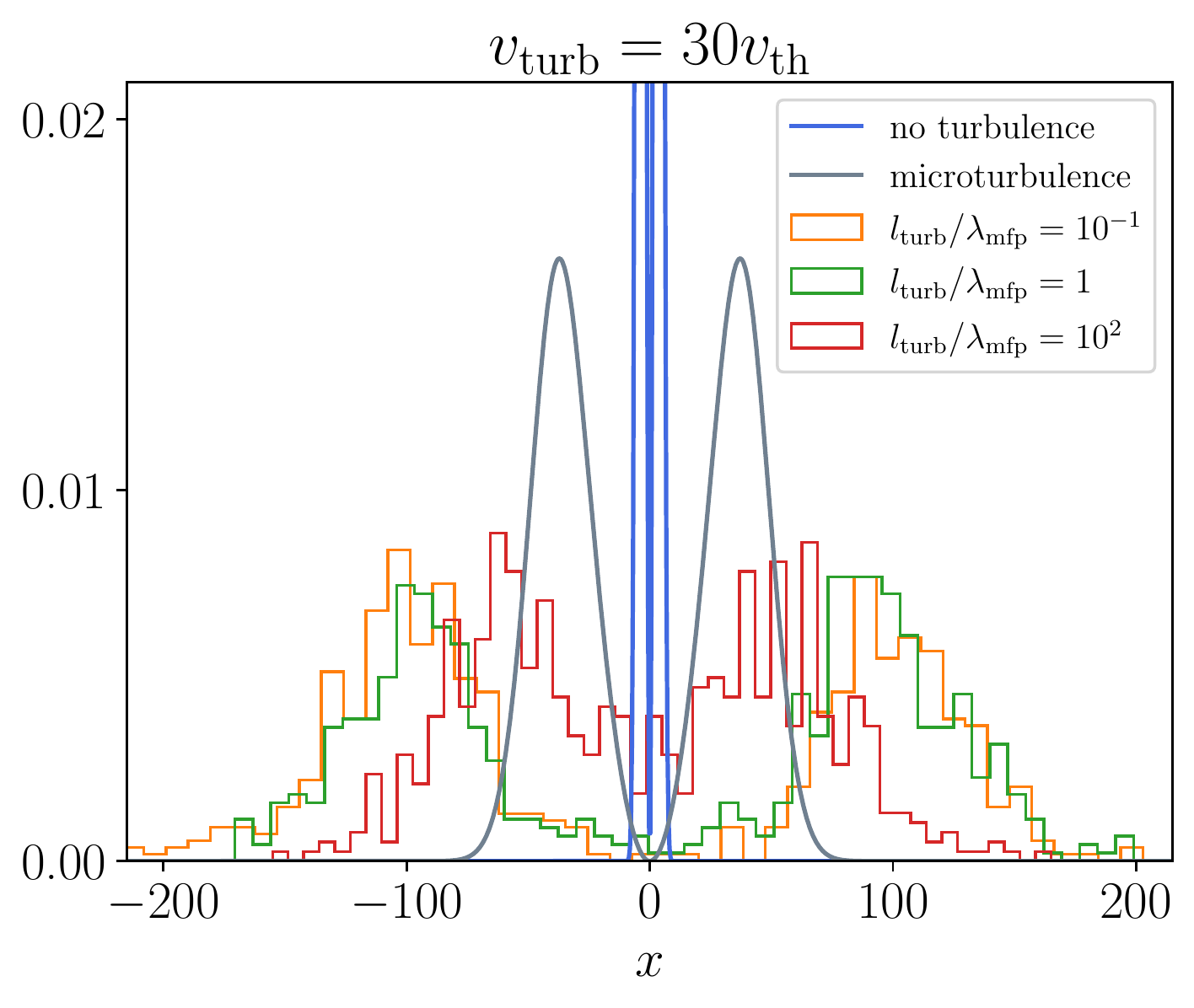}

\caption{\label{fig_spectrum}The spectrum of Ly$\alpha$ photons emerging
from a turbulent sphere with the line center optical depth of $\tau_{0}=10^{5}$
and temperature of $T=10^{4}\:\textrm{K}$. For each fixed value of
the turbulence velocity $v_{\textrm{turb}}$ {[}$v_{\textrm{turb}}=0.1v_{\textrm{th}}$
(top left), $v_{\textrm{turb}}=v_{\textrm{th}}$ (top right), $v_{\textrm{turb}}=10v_{\textrm{th}}$
(bottom left), $v_{\textrm{turb}}=30v_{\textrm{th}}$ (bottom right){]}
each subplot shows the histogram of the emergent spectrum for several
values of the turbulence correlation length $l_{\textrm{turb}}$ {[}$l_{\textrm{turb}}=10^{-1}\lambda_{\textrm{mfp}}$
(orange), $l_{\textrm{turb}}=\lambda_{\textrm{mfp}}$ (green), $l_{\textrm{turb}}=10^{2}\lambda_{\textrm{mfp}}$
(red){]} as well as the Neufeld analytical solution \citep{Neufeld1990}
for the cases with no turbulence (royalblue) and with microturbulence
(slategray).}
\end{figure*}

\section{Model description}

\label{Section_2}

\begin{figure}
\includegraphics[width=1\columnwidth]{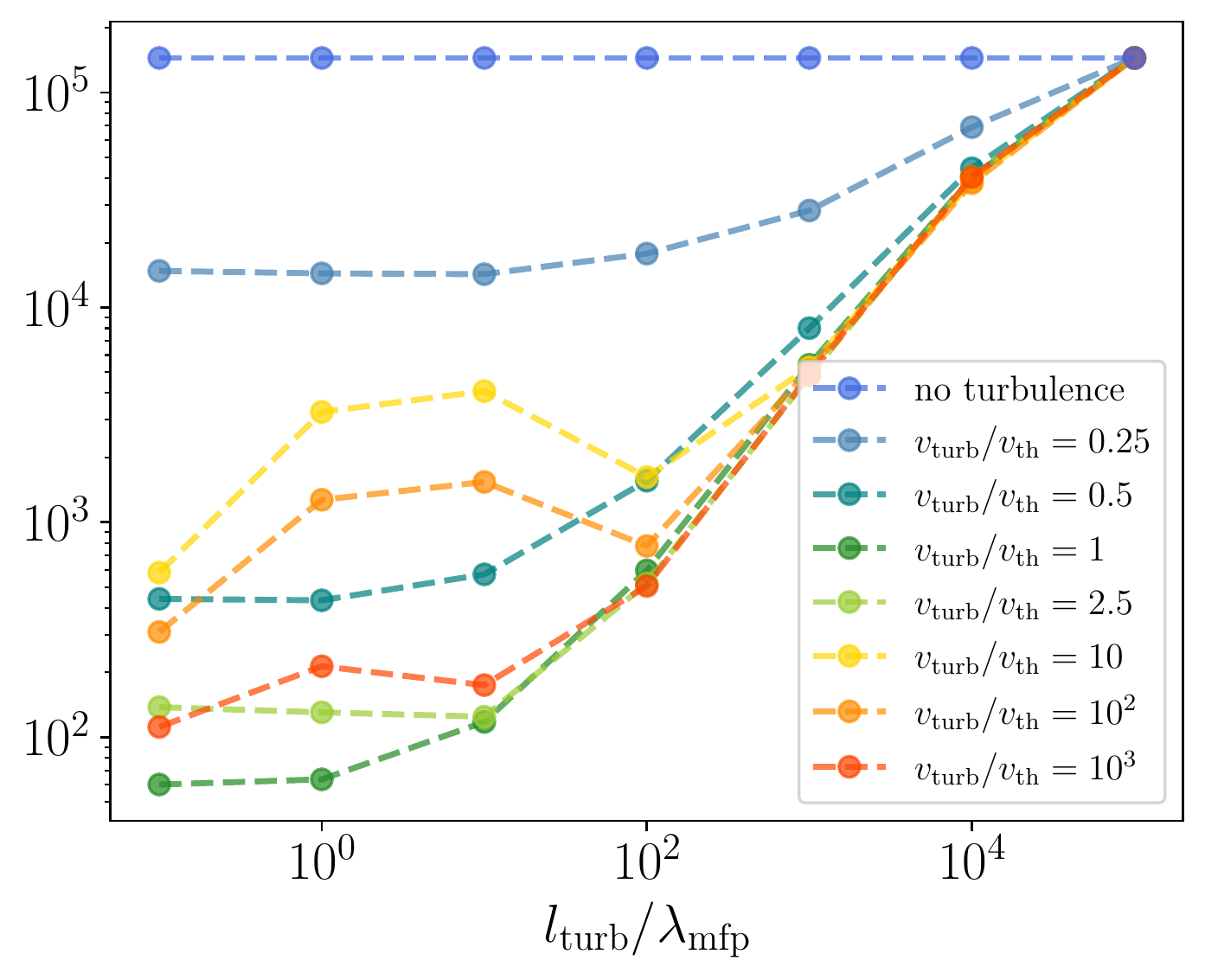}

\includegraphics[width=1\columnwidth]{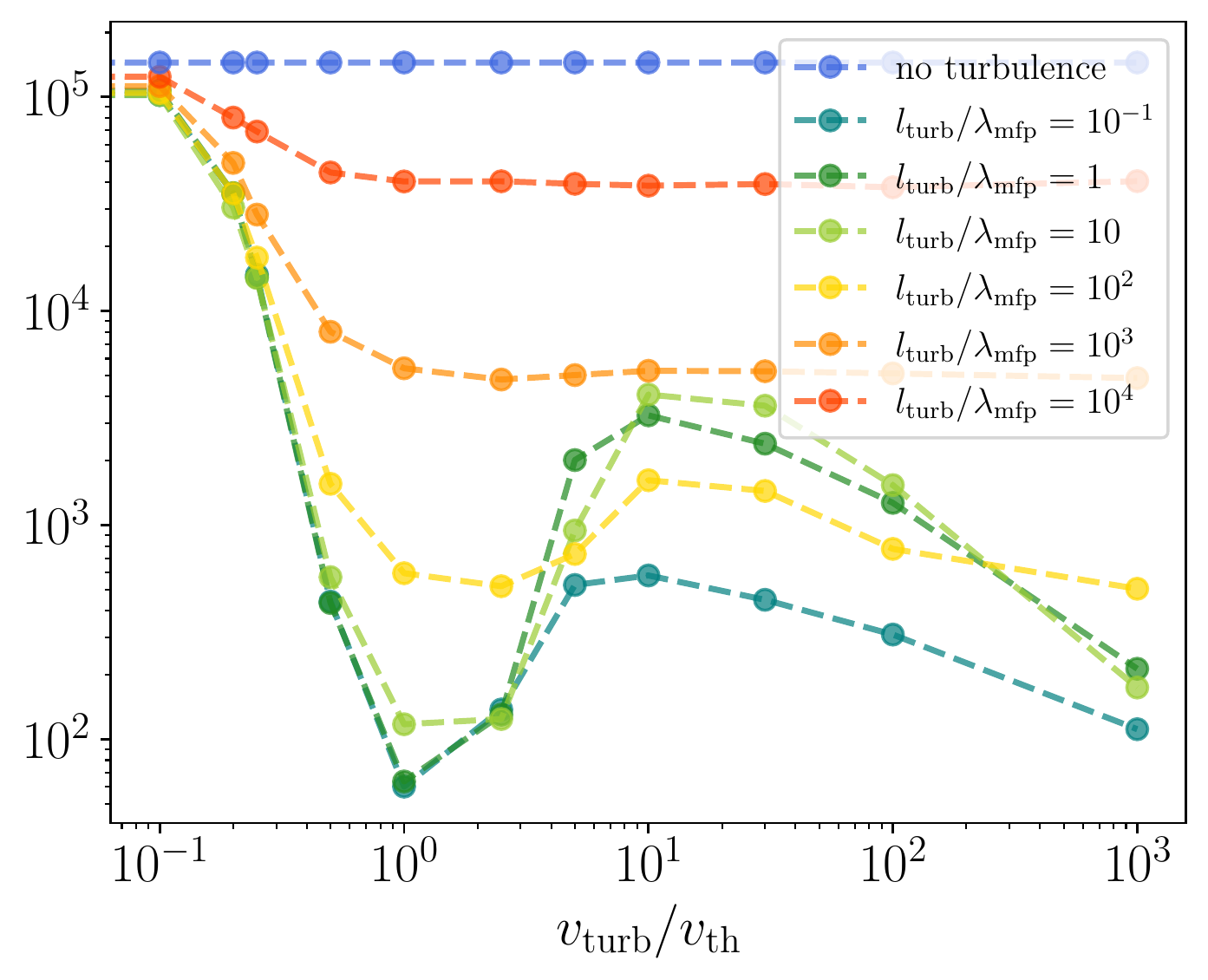}

\caption{\label{Nscat_vs_lturb_vturb}Top panel: The average number of scattering
events $N_{\textrm{scat}}$ that a representative photon undergoes
before it escapes the turbulent cloud of neutral hydrogen versus the
turbulence correlation length $l_{\textrm{turb}}$ for different values
of the turbulence velocity amplitude $v_{\textrm{turb}}$ {[}$v_{\textrm{turb}}=0.25v_{\textrm{th}}$
(steelblue), $v_{\textrm{turb}}=0.5v_{\textrm{th}}$ (teal), $v_{\textrm{turb}}=v_{\textrm{th}}$
(forestgreen), $v_{\textrm{turb}}=2.5v_{\textrm{th}}$ (yellowgreen),
$v_{\textrm{turb}}=10v_{\textrm{th}}$ (gold), $v_{\textrm{turb}}=10^{2}v_{\textrm{th}}$
(darkorange), $v_{\textrm{turb}}=10^{3}v_{\textrm{th}}$ (orangered){]}
as well as for the case with no turbulence (royalblue). Bottom panel:
The average number of scattering events $N_{\textrm{scat}}$ that
a representative photon undergoes versus the turbulence velocity $v_{\textrm{turb}}$
for different values of the turbulence correlation length $l_{\textrm{turb}}$
{[}$l_{\textrm{turb}}=10^{-1}\lambda_{\textrm{mfp}}$ (teal), $l_{\textrm{turb}}=\lambda_{\textrm{mfp}}$
(forestgreen), $l_{\textrm{turb}}=10\lambda_{\textrm{mfp}}$ (yellowgreen),
$l_{\textrm{turb}}=10^{2}\lambda_{\textrm{mfp}}$ (gold), $l_{\textrm{turb}}=10^{3}\lambda_{\textrm{mfp}}$
(darkorange), $l_{\textrm{turb}}=10^{4}\lambda_{\textrm{mfp}}$ (orangered){]}
as well as for the case with no turbulence (royalblue).}
\end{figure}

Our Monte Carlo radiative transfer code, except for the turbulence
handling, largely follows the Monte Carlo procedure described in~\citet{Dijkstra2017}
(see also \citealt{Laursen2010}). Our goal is to have a macroscopic (bulk) velocity field $\mathbf{v}_{\textrm{turb}}$ that has an effective correlation length $l_{\textrm{turb}}$. There could be at least two ways to accomplish this. First approach is to divide the sphere into the cells of size $l_{\textrm{turb}}$ with each cell having its own $\mathbf{v}_{\textrm{turb}}$ drawn from the normal distribution and then launch photons through this fixed preassigned grid. However, such a procedure would be unnecessary memory and computationally heavy for our purposes, since we are interested in the averaged properties of the escaped radiation rather than the precise distribution inside the sphere. So instead we pursue the following approach. We draw a new random velocity $\mathbf{v}_{\textrm{turb}}$ from the normal distribution each time photon traveled a physical distance exceeding $l_{\textrm{turb}}$. Such division into "cells" is not preassigned but generated on the fly for each photon independently as it propagates. We note that some similar ideas were also used in \citet{Magnan1976} and labeled as the effective cells approach.

First, we run our code without turbulence and make sure it agrees
with known analytical solutions or previously studied numerical models.
In particular we compared the solutions for static, expanding, and
contracting uniform spherical cloud such as those studied in~\citet{Zheng2002}.
We also considered three models of anisotropic hydrogen clouds, namely
the density gradient, the velocity gradient, and the bipolar wind
models, as described in~\citet{Zheng2014}. The code, its description,
and the test results can be found in the GitHub repository~\citet{Munirov_LyAMC_2022} (and in Appendix here).

Throughout the paper we use definitions of the variables and functions
as those in~\citet{Dijkstra2017}, unless stated otherwise explicitly.
We assume that the readers are familiar with the basics of the Ly$\alpha$
transfer problem, otherwise it is recommended to consult~\citet{Dijkstra2017}
(see also \citealt{Laursen2010}). Specifically, we use the dimensionless
photon frequency $x=\left(\nu-\nu_{\alpha}\right)/\Delta\nu_{\alpha}$,
where $\Delta\nu_{\alpha}/\nu_{\alpha}=b/c$, $b$ is the Doppler
broadening parameter, $c$ is the speed of light, and $\nu_{\alpha}=2.47\times10^{15}\:\textrm{Hz}$
is the frequency at the Ly$\alpha$ line center. We use $\lambda_{\textrm{mfp}}$
to denote the mean free path of a photon at the line center $\lambda_{\textrm{mfp}}\equiv\lambda_{\textrm{mfp}}\left(x=0\right)$
in the hydrogen cloud without turbulence. We denote the optical depth
from the center of the cloud to the edge for the line center photons
without turbulence by $\tau_{0}$. We define the thermal velocity
as $v_{\textrm{th}}=\sqrt{2k_{B}T/m_{p}}$, where $k_{B}$ is the
Boltzmann constant, $m_{p}$ is the proton mass, and $T$ is the gas
temperature in Kelvins. We use the following definition for the Voigt
function $H(a,x)$:

\noindent 
\begin{equation}
H(a,x)=\frac{a}{\pi}\int_{-\infty}^{\infty}\frac{e^{-\xi^{2}}d\xi}{\left(\xi-x\right)^{2}+a^{2}},
\end{equation}

\noindent where $a=4.7\times10^{-4}$$(T/10^{4}\:\textrm{K})^{-1/2}$
is the Voigt parameter. In all our simulations of the turbulence we
consider a uniform sphere of radius $R=1$ (so that all the distances
are measured in the units of $R$,  in particular the mean free path at the line center is $\lambda_{\textrm{mfp}}=\tau_{0}^{-1}$) with a point source at the center
of the sphere; we include the recoil effect but ignore photon absorption
due to dust; we consider Ly$\alpha$ system decoupled from a Hubble
flow. The integration step is chosen to be a small enough fraction
of $l_{\textrm{turb}}$, so that we always resolve a turbulent cell
numerically.

In the next section we present and discuss the numerical solutions
obtained using the Monte Carlo code described in this section.

\begin{figure*}
\includegraphics[width=0.5\textwidth]{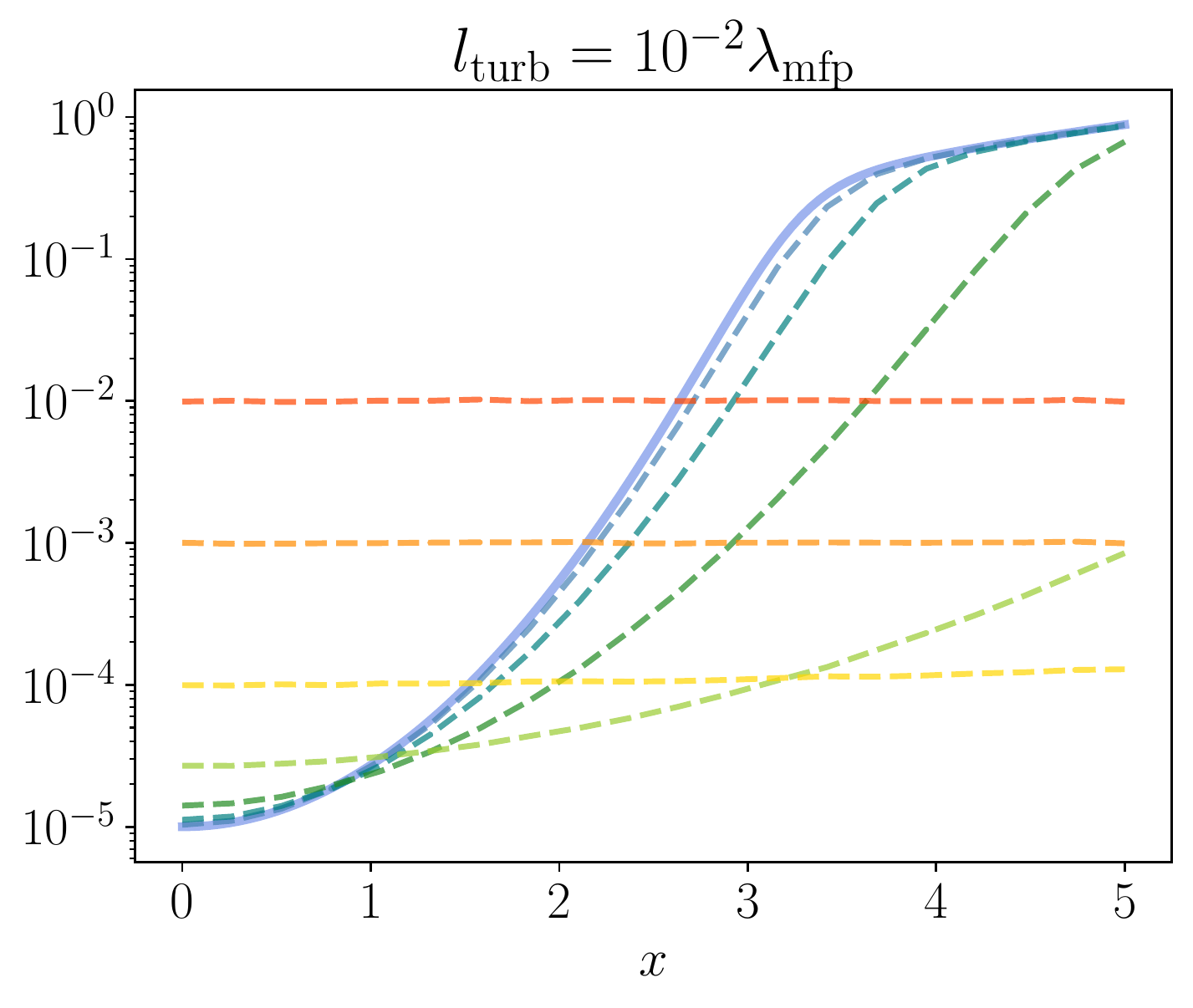}\includegraphics[width=0.5\textwidth]{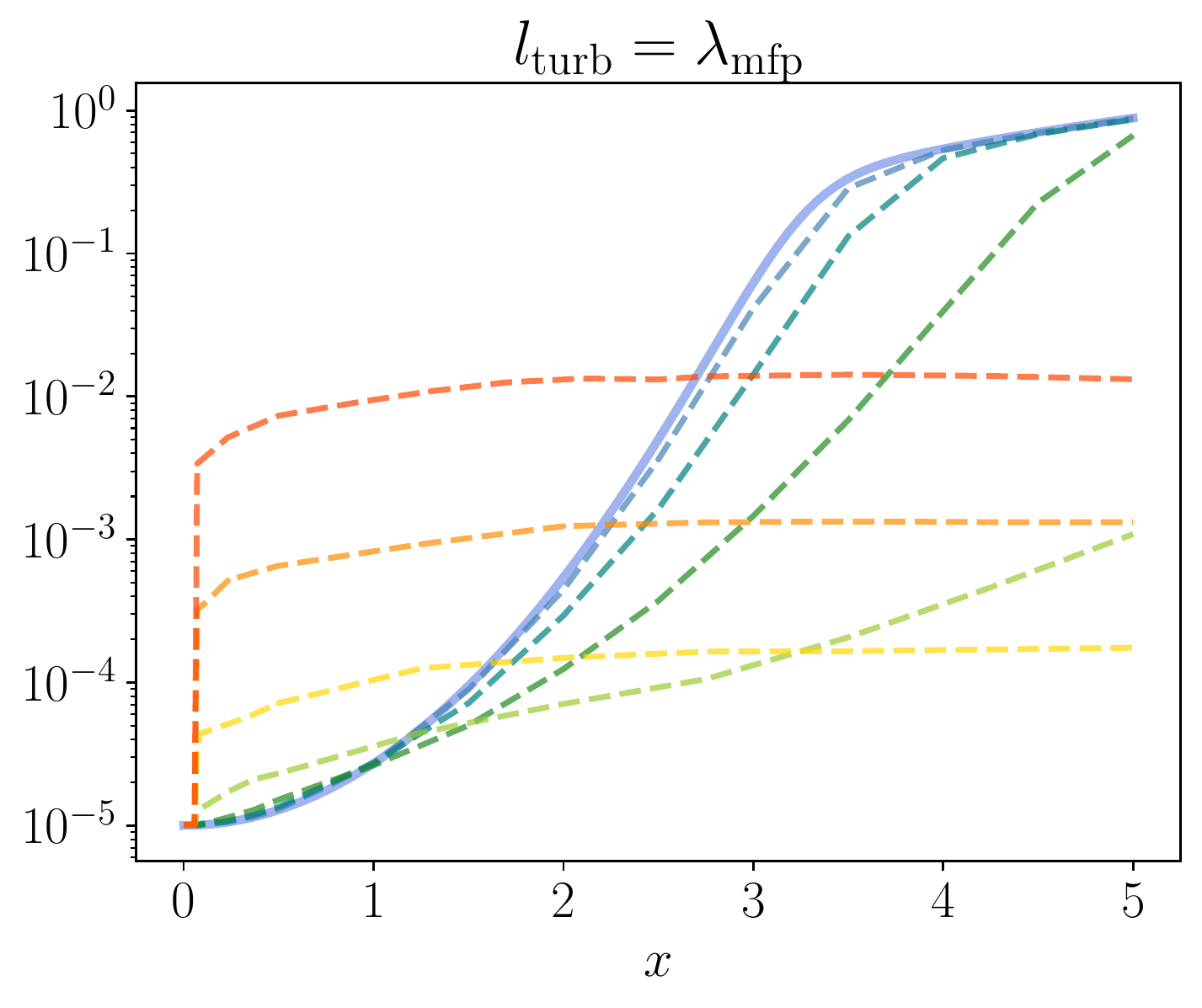}

\includegraphics[width=0.5\textwidth]{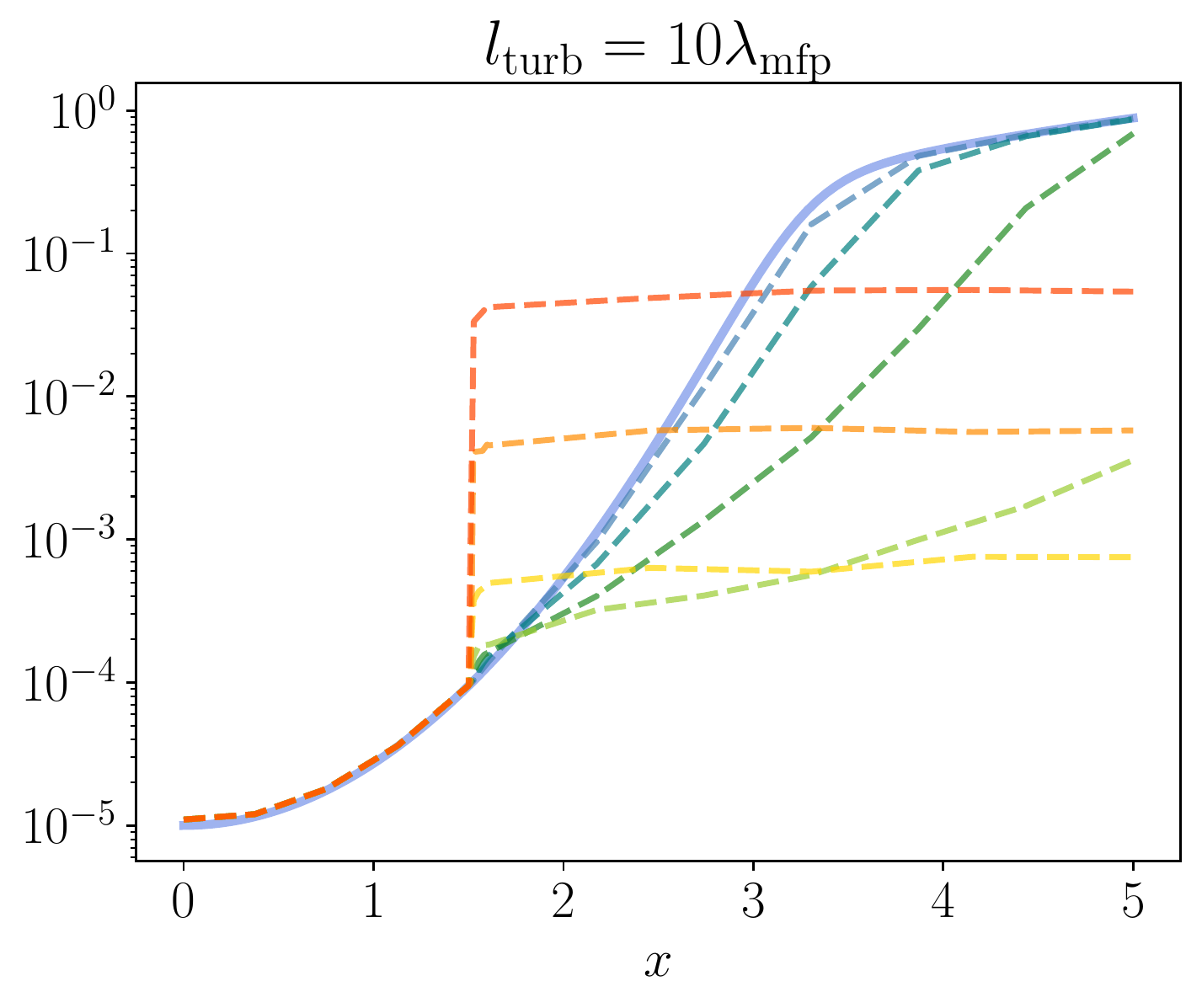}\includegraphics[width=0.5\textwidth]{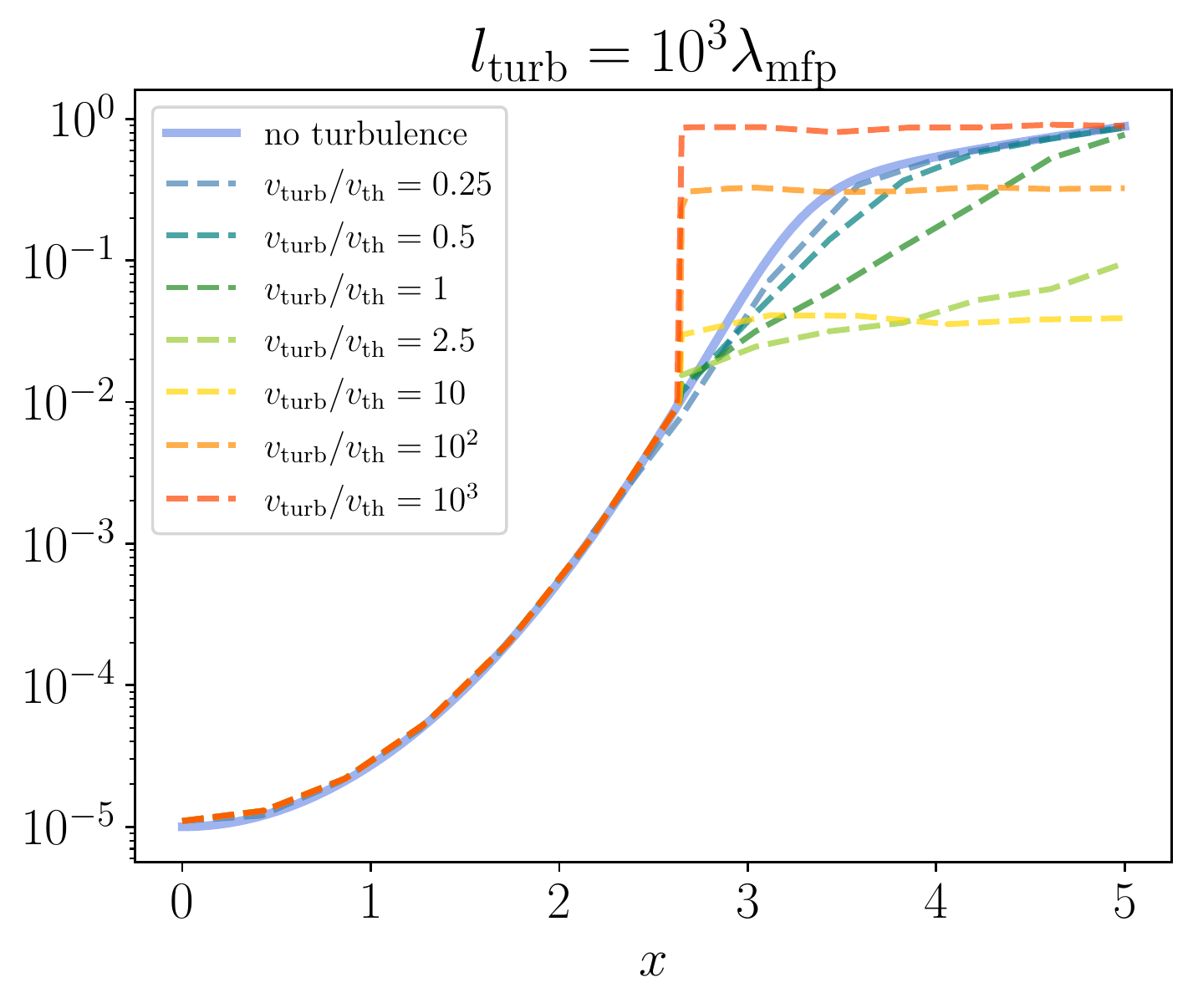}

\caption{\label{mfp_fixed_lturb}The numerically calculated effective mean
free path versus the dimensionless frequency $x$. For each fixed
value of the turbulence correlation length $l_{\textrm{turb}}$ {[}$l_{\textrm{turb}}=10^{-2}\lambda_{\textrm{mfp}}$
(top left), $l_{\textrm{turb}}=\lambda_{\textrm{mfp}}$ (top right),
$l_{\textrm{turb}}=10\lambda_{\textrm{mfp}}$ (bottom left), $l_{\textrm{turb}}=10^{3}\lambda_{\textrm{mfp}}$
(bottom right){]} the corresponding subplot shows the effective mean
three path as a function of $x$ for several values of the turbulence
velocity $v_{\textrm{turb}}$ {[}$v_{\textrm{turb}}=0.25v_{\textrm{th}}$
(steelblue), $v_{\textrm{turb}}=0.5v_{\textrm{th}}$ (teal), $v_{\textrm{turb}}=v_{\textrm{th}}$
(forestgreen), $v_{\textrm{turb}}=2.5v_{\textrm{th}}$ (yellowgreen),
$v_{\textrm{turb}}=10v_{\textrm{th}}$ (gold), $v_{\textrm{turb}}=10^{2}v_{\textrm{th}}$
(darkorange), $v_{\textrm{turb}}=10^{3}v_{\textrm{th}}$ (orangered){]}
as well as for the case with no turbulence (royalblue).}
\end{figure*}

\begin{figure*}
\includegraphics[width=0.5\textwidth]{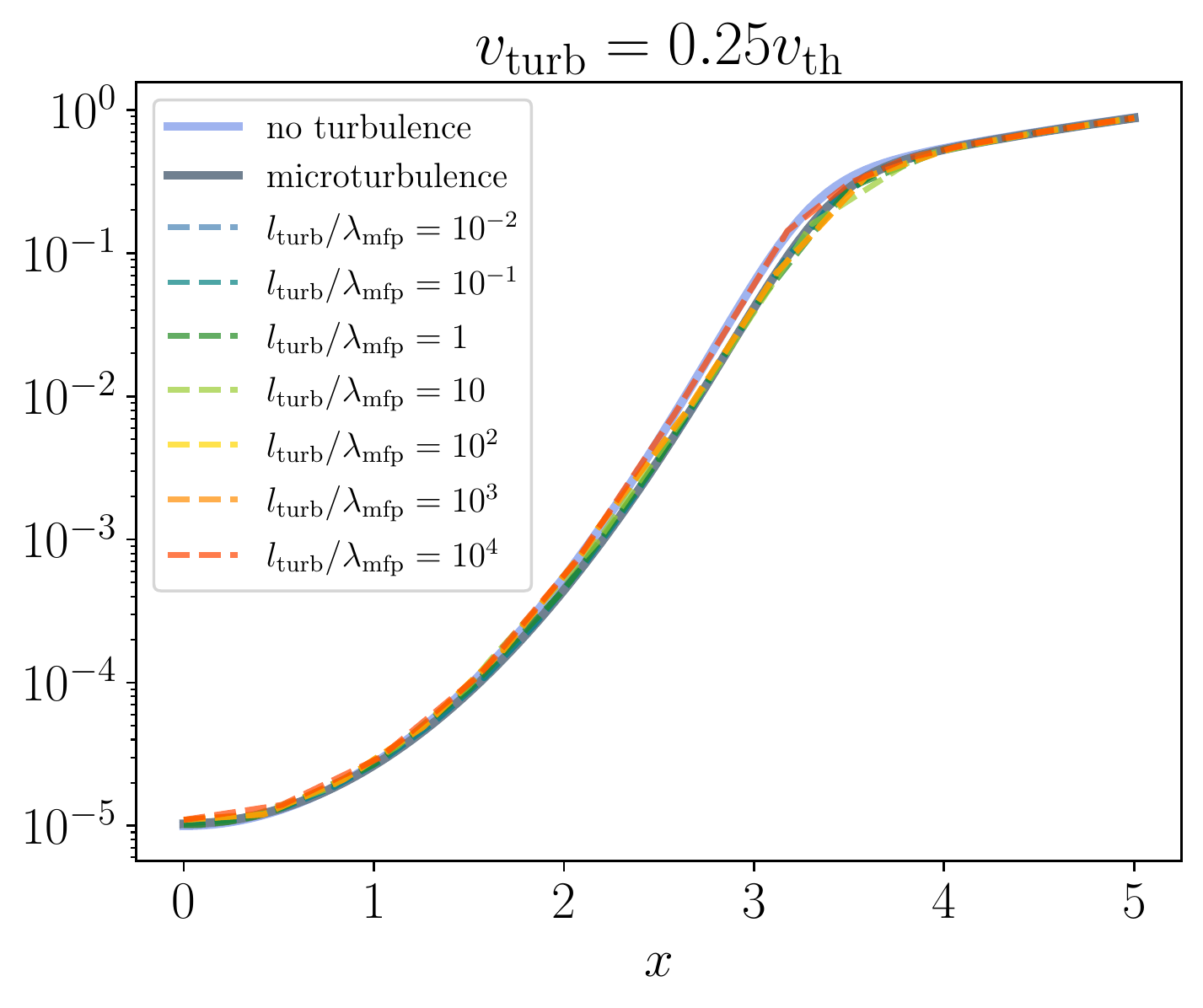}\includegraphics[width=0.5\textwidth]{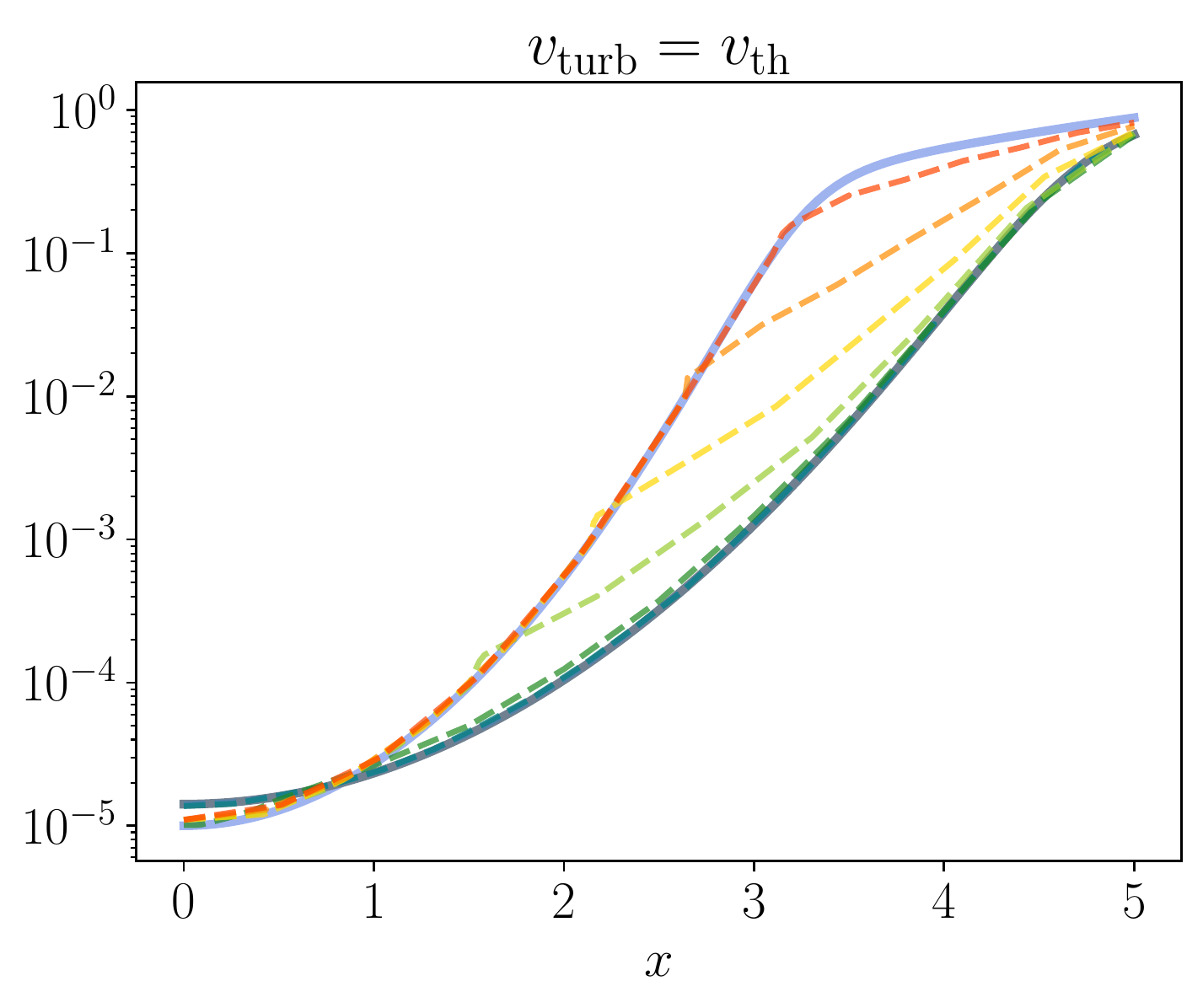}

\includegraphics[width=0.5\textwidth]{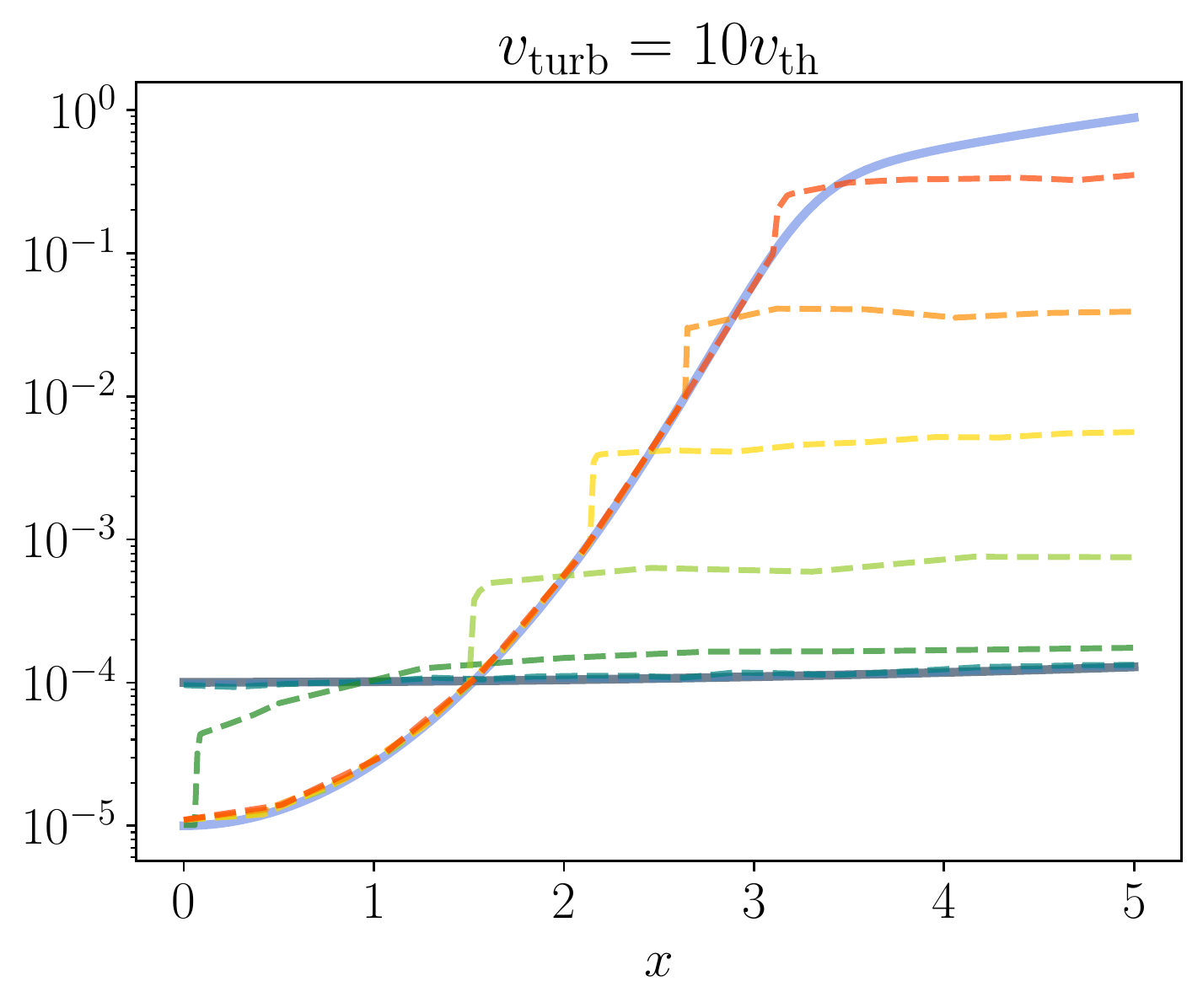}\includegraphics[width=0.5\textwidth]{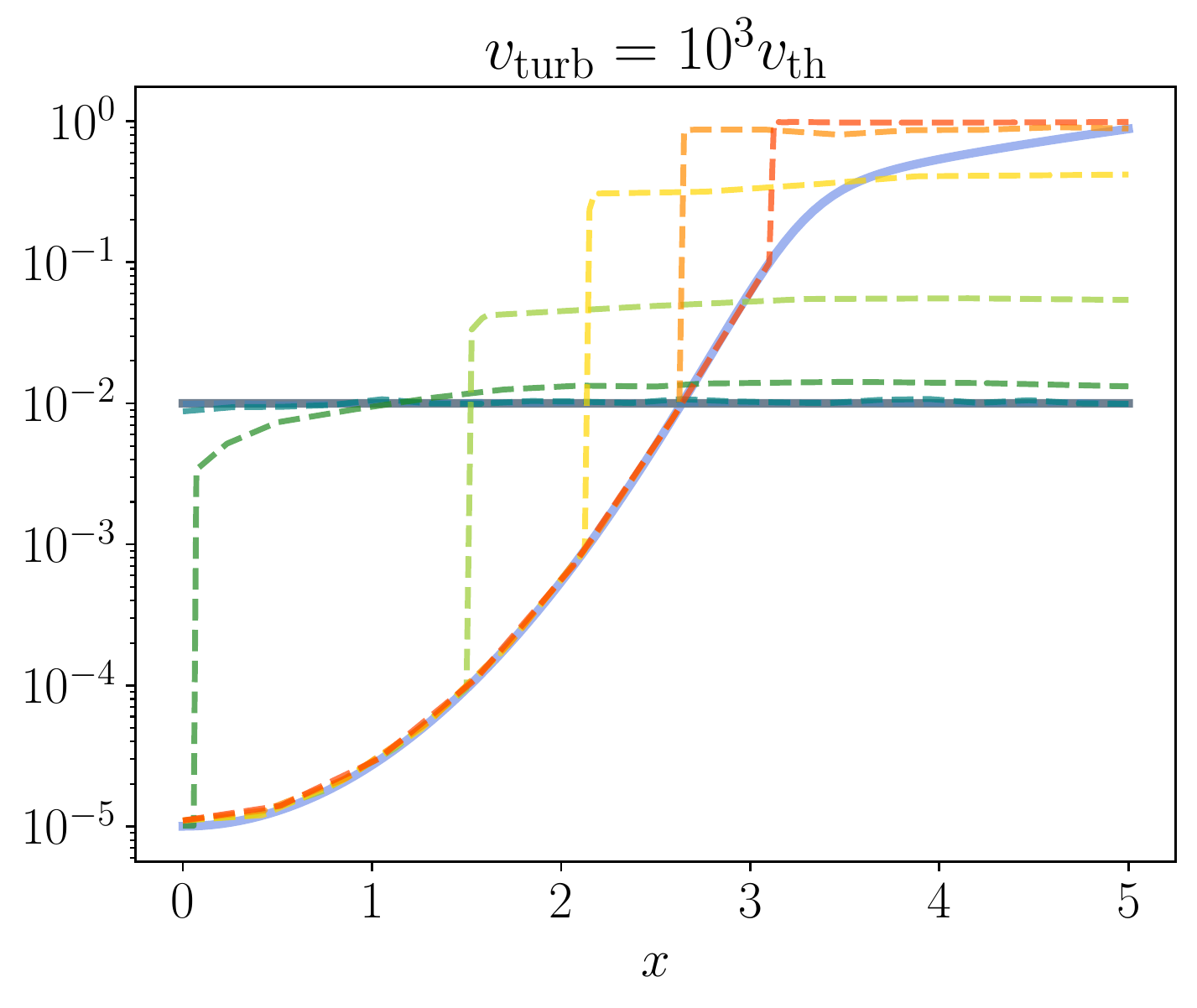}

\caption{\label{mfp_fixed_vturb}The numerically calculated effective mean
free path versus the dimensionless frequency $x$. For each fixed
value of the turbulence velocity $v_{\textrm{turb}}$ {[}$v_{\textrm{turb}}=0.25v_{\textrm{th}}$
(top left), $v_{\textrm{turb}}=v_{\textrm{th}}$ (top right), $v_{\textrm{turb}}=10v_{\textrm{th}}$
(bottom left), $v_{\textrm{turb}}=10^{3}v_{\textrm{th}}$ (bottom
right){]} the corresponding subplot shows the effective mean three
path as a function of $x$ for the following values of the turbulence
correlation length $l_{\textrm{turb}}$ {[}$l_{\textrm{turb}}=10^{-2}\lambda_{\textrm{mfp}}$
(steelblue), $l_{\textrm{turb}}=10^{-1}\lambda_{\textrm{mfp}}$ (teal),
$l_{\textrm{turb}}=\lambda_{\textrm{mfp}}$ (forestgreen), $l_{\textrm{turb}}=10\lambda_{\textrm{mfp}}$
(yellowgreen), $l_{\textrm{turb}}=10^{2}\lambda_{\textrm{mfp}}$ (gold),
$l_{\textrm{turb}}=10^{3}\lambda_{\textrm{mfp}}$ (darkorange), $l_{\textrm{turb}}=10^{4}\lambda_{\textrm{mfp}}$
(orangered){]} as well as for the cases with no turbulence (royalblue)
and with microturbulence (slategray).}
\end{figure*}

\section{Simulations results}

\label{Section_3}

Let us consider a uniform turbulent cloud of neutral hydrogen with
the line center optical depth of $\tau_{0}=10^{5}$ and with the typical
temperature of the neutral gas in the intergalactic medium of $T=10^{4}\:\textrm{K}$.
This corresponds to the hydrogen column density of $n_{\textrm{HI}}\approx1.7\times10^{18}\:\textrm{c\ensuremath{\textrm{m}^{-2}}}$
and the Voigt parameter $a$ such that $a\tau_{0}\approx47$.

We run the Monte Carlo simulations for several values of the turbulence
correlation length $l_{\textrm{turb}}$ ranging from $l_{\textrm{turb}}=10^{-1}\lambda_{\textrm{mfp}}$
to $l_{\textrm{turb}}=10^{4}\lambda_{\textrm{mfp}}$ and several values
of the turbulence velocity amplitude $v_{\textrm{turb}}$ ranging
from $v_{\textrm{turb}}=0$ to $v_{\textrm{turb}}=10^{3}v_{\textrm{th}}$.
We launch $10^{5}$ photons for each set of the parameters and when
the photons leave the turbulent cloud we register their frequencies
and the total number of scatterings they experienced.

The spectrum of the Ly$\alpha$ photons emerging from a spherical
cloud of turbulent neutral hydrogen is shown in Fig.~\ref{fig_spectrum}.
For a fixed value of the turbulence velocity $v_{\textrm{turb}}$
($v_{\textrm{turb}}/v_{\textrm{th}}=0.1,1,10,30$) each subplot shows
the histogram of the emerging spectrum for three values of the turbulence
correlation length $l_{\textrm{turb}}$ ($l_{\textrm{turb}}/\lambda_{\textrm{mfp}}=10^{-1},1,10^{2}$)
as well as the Neufeld analytical solution \citep{Neufeld1990} for
the cases without turbulence and with microturbulence for the corresponding
value of $v_{\textrm{turb}}$.

We can see from Fig.~\ref{fig_spectrum} that in the presence of
turbulence the emergent spectrum is still double peaked but the peaks
of the spectrum move further away from the line center. Due to the
recoil effect, the spectrum has a slight asymmetry with more photons
scattered into smaller frequencies, but for larger values of the turbulence
velocity, this asymmetry diminishes. For $v_{\textrm{turb}}\lesssim v_{\textrm{th}}$,
the emergent spectrum can be approximately described by the microturbulence
model, while for $v_{\textrm{turb}}\gtrsim v_{\textrm{th}}$, the
difference is substantial even for small correlation lengths of the
turbulence ($l_{\textrm{turb}}=10^{-1}\lambda_{\textrm{mfp}}$).

The more surprising result is a significant reduction in the total
number of scattering events $N_{\textrm{scat}}$ undergone by the
Ly$\alpha$ photons before they escape the gas cloud. The top row
of Fig.~\ref{Nscat_vs_lturb_vturb} shows the average number of scatterings
that a Ly$\alpha$ photon experiences before it escapes as a function
of the turbulence correlation length $l_{\textrm{turb}}$ for different
values of the turbulence velocity amplitude $v_{\textrm{turb}}$,
while the bottom row of Fig.~\ref{Nscat_vs_lturb_vturb} shows the
average number of scatterings $N_{\textrm{scat}}$ versus the turbulence
velocity $v_{\textrm{turb}}$ for different values of the turbulence
correlation length $l_{\textrm{turb}}$. In addition, the number of
scatterings events $N_{\textrm{scat}}$ for the case without turbulence
is shown in both subplots.

We can see from Fig.~\ref{Nscat_vs_lturb_vturb} that without turbulence
the average number of scattering events $N_{\textrm{scat}}$ is at
the order of the optical depth $\tau_{0}=10^{5}$, which agrees with
the well known result \citep{Adams1972,Harrington1973,Dijkstra2017}
that in the absence of turbulence the number of scatterings for optically
thick case is given by $N_{\textrm{scat}}=C\tau_{0}$, where $C$
is a numerical constant of order one. We can also see that for $v_{\textrm{turb}}/v_{\textrm{th}}\rightarrow0$
and $l_{\textrm{turb}}/\lambda_{\textrm{mfp}}\rightarrow\infty$ the
average number of scatterings is also approaching $\tau_{0}=10^{5}$,
as expected because these cases converge to the absence of turbulence.
Overall, the number of scatterings depends on $l_{\textrm{turb}}$
and $v_{\textrm{turb}}$ in a complicated manner, but, similarly to
the case of expanding/contracting gas \citep{Bonilha1979}, there
is a decrease in the number of scattering events for all parameters
of turbulence.

Turbulence influences the scattering process in two ways. First, for
a given optical depth generated from an exponential distribution $P\left(\tau\right)=e^{-\tau}$
the effective mean free path is changed because the photon frequency
in the local frame of the gas depends on the turbulent bulk motion
of the medium: $x^{\prime}=x-\mathbf{v}_{\textrm{bulk}}\left(\mathbf{r}^{\prime}\right)\cdot\mathbf{k}/v_{\textrm{th}}$.
Second, once the scattering cell is determined, the photon besides
experiencing the usual frequency redistribution in the gas frame,
gets an additional Lorentz transformed frequency shift $\mathbf{v}_{\textrm{bulk}}\cdot\left(\mathbf{k}_{\textrm{out}}-\mathbf{k}_{\textrm{in}}\right)/v_{\textrm{th}}$
in the laboratory frame. This leads to an additional random jump in
frequency space on the order of $\left|\triangle x\right|\approx v_{\textrm{turb}}/v_{\textrm{th}}$
and results in faster transition from the core to the wings of the
Voigt profile. Since scattering in the cell is always local by definition,
once the scattering cell is determined the turbulence correlation
length does not matter. Thus, even if the correlation length is very
small, turbulence can still influence the photon propagation through
the above mentioned random frequency jumps.

To better understand how the presence of turbulence with a finite
correlation length changes the mean free path for propagating photons,
we computed the effective mean free path in the following way: for
an optical depth drawn from a Poisson distribution with $\tau=1$
a test photon of frequency $x$ is launched a sufficient number of
times (hundreds was enough to achieve stable results) with fixed values
of $v_{\textrm{turb}}$ and $l_{\textrm{turb}}$ and the effective
mean free path is then calculated as an average of these launches. 

Figs.~\ref{mfp_fixed_lturb}~and~\ref{mfp_fixed_vturb} show the
effective mean free path versus the dimensionless frequency $x$.
In Fig.~\ref{mfp_fixed_lturb}, for each fixed value of the turbulence
correlation length $l_{\textrm{turb}}$ ($l_{\textrm{turb}}/\lambda_{\textrm{mfp}}=10^{-2},1,10,10^{3}$)
we plot the effective mean free path as a function of $x$ for several
values of the turbulence velocity $v_{\textrm{turb}}$ ($v_{\textrm{turb}}/v_{\textrm{th}}=0.25,0.5,1,2.5,10,10^{2},10^{3}$)
and for the case with no turbulence. In Fig.~\ref{mfp_fixed_vturb},
for each fixed value of the turbulence velocity $v_{\textrm{turb}}$
($v_{\textrm{turb}}/v_{\textrm{th}}=0.25,1,10,10^{3}$) we plot the
effective mean free path as a function of $x$ for several values
of the turbulence correlation length $l_{\textrm{turb}}$ ($l_{\textrm{turb}}/\lambda_{\textrm{mfp}}=10^{-2},10^{-1},1,10,10^{2},10^{3},10^{4}$)
as well as the mean free paths in the absence of turbulence and with
microturbulence for the corresponding value of $v_{\textrm{turb}}$.
In the absence of turbulence, one gets the usual mean free path determined
by the inverse of the Voigt profile: $\lambda_{\textrm{mfp}}\left(x\right)=1/\tau_{0}H\left(a,x\right)$,
while for microturbulence one gets a properly scaled inverse of the
Voigt profile: 
\begin{equation}
\lambda_{\textrm{mfp}}\left(x\right)=\frac{\sqrt{1+v_{\textrm{turb}}^{2}/v_{\textrm{th}}^{2}}}{\tau_{0}H\left(\frac{a}{\sqrt{1+v_{\textrm{turb}}^{2}/v_{\textrm{th}}^{2}}},\frac{x}{\sqrt{1+v_{\textrm{turb}}^{2}/v_{\textrm{th}}^{2}}}\right)}.
\end{equation}

We see from Figs.~\ref{mfp_fixed_lturb}~and~\ref{mfp_fixed_vturb}
that for $l_{\textrm{turb}}\ll\lambda_{\textrm{mfp}}$ ($l_{\textrm{turb}}/\lambda_{\textrm{mfp}}=10^{-2},10^{-1}$),
the effective mean free path indeed matches the microturbulence curve
for all values of the turbulence velocity. On the other hand, for
large values of the turbulence correlation length $l_{\textrm{turb}}\gtrsim\lambda_{\textrm{mfp}}$,
the effective mean free path follows the mean free path $\lambda_{\textrm{mfp}}\left(x\right)=1/\tau_{0}H\left(a,x\right)$
in the absence of turbulence up until some threshold value $x_{l}$,
where the effective mean free path experiences jump and changes its
dependence. The value of the threshold is essentially independent
of the turbulence velocity amplitude $v_{\textrm{turb}}$ and depends
mainly on the turbulence correlation length $l_{\textrm{turb}}$,
and is approximately determined by the condition $\lambda_{\textrm{mfp}}\left(x_{l}\right)=l_{\textrm{turb}}$
or, equivalently, $H\left(a,x_{l}\right)=l_{\textrm{turb}}^{-1}$.

Going back to Fig.~\ref{Nscat_vs_lturb_vturb}, we see that for $v_{\textrm{turb}}\lesssim v_{\textrm{th}}$,
the number of scatterings $N_{\textrm{scat}}$ decreases as the turbulence
velocity amplitude $v_{\textrm{turb}}$ increases. It can be understood
in the following way. While the effective mean free path is not changed
dramatically for $v_{\textrm{turb}}\lesssim v_{\textrm{th}}$ (see
Fig.~\ref{mfp_fixed_vturb}), turbulence still brings random jumps
in frequency space with the characteristic size $\left|\triangle x\right|\approx v_{\textrm{turb}}/v_{\textrm{th}}\lesssim1$;
these random jumps allow photons to move from the core to the wings
in fewer scatterings (the core-wing boundary for our parameters is
at $\left|x\right|\approx3.26$). The bigger those jumps, the faster
the photons jump from the core to the wings, which is why we see a decrease
in the number of scatterings for $v_{\textrm{turb}}\lesssim v_{\textrm{th}}$
as $v_{\textrm{turb}}$ approaches $v_{\textrm{th}}$. For larger
values of the turbulence velocity amplitude $v_{\textrm{turb}}\gtrsim v_{\textrm{th}}$,
we see that the number of scatterings $N_{\textrm{scat}}$ generally
exhibits a non-monotonic dependence on $v_{\textrm{turb}}$. Here
we have a competition between the change in the effective mean free
path and large, often exceeding the width of the core region, random
frequency jumps on the order of $\left|\triangle x\right|\approx v_{\textrm{turb}}/v_{\textrm{th}}\gtrsim1$.

We also see from Fig.~\ref{Nscat_vs_lturb_vturb} that the number
of scatterings $N_{\textrm{scat}}$ generally speaking decreases as
the turbulence correlation length $l_{\textrm{turb}}$ decreases.
At small but finite values of the turbulence correlation length $l_{\textrm{turb}}$
the random frequency jumps are largely localized in space and happen
often enough to drive Ly$\alpha$ photons to the wings, which allows
them to escape in fewer scatterings. For sufficiently large values
of $l_{\textrm{turb}}$, the number of scatterings inside a region
of size $l_{\textrm{turb}}$ is approximately given by $\tau_{\textrm{eff}}=l_{\textrm{turb}}/\lambda_{\textrm{mfp}}$,
if in addition $v_{\textrm{turb}}\gtrsim v_{\textrm{th}}$, then once
the photons random walked out of the region of size $l_{\textrm{turb}}$
it will likely escape, so that the total number of scatterings is
approximately $N_{\textrm{scat}}\approx\tau_{\textrm{eff}}=l_{\textrm{turb}}/\lambda_{\textrm{mfp}}$.
Indeed, we see from Fig.~\ref{Nscat_vs_lturb_vturb} that for $v_{\textrm{turb}}\gtrsim v_{\textrm{th}}$
and for large values of the turbulence correlation length between
approximately $l_{\textrm{turb}}/\lambda_{\textrm{mfp}}=10$ and $l_{\textrm{turb}}/\lambda_{\textrm{mfp}}=10^{5}$
the number of scatterings scales as $N_{\textrm{scat}}\approx4l_{\textrm{turb}}/\lambda_{\textrm{mfp}}$.

We conclude by pointing out that the exact dependence of the number
of scatterings $N_{\textrm{scat}}$ on $l_{\textrm{turb}}$ and $v_{\textrm{turb}}$
and its sensitivity to changes in these parameters is complicated
and also depends on the characteristics, such as density and temperature,
of the $\textrm{HI}$ regions though which Ly$\alpha$ photons propagate\footnote{In the GitHub repository~\citet{Munirov_LyAMC_2022}  (and in Appendix here) we provide plots for $\tau_{0}=10^{5}$
and $T=1\:\textrm{K}$, which corresponds to a strongly optically
thick case with the Voigt parameter $a$ such that $a\tau_{0}\approx4.7\times10^{3}$.}. However, overall, the presence of turbulence makes it easier for
Ly$\alpha$ photons to be scattered into the wings, facilitating their
escape and reducing the number of scattering events they experience.

\section{Conclusion}

\label{Section_4}

In this paper we focused on the qualitative analysis of the turbulence
effect on the Ly$\alpha$ transfer. We adopted a Monte Carlo approach
and considered a simple geometry with turbulence represented as spatial
domains of given size with randomly directed velocities.

We performed numerical simulations and discovered that the presence
of turbulence not only alters the emergent spectrum, but turbulence
with small but finite correlation length can significantly, by orders
of magnitude, reduce the number of scatterings required for the Ly$\alpha$
photons to escape the cloud of neutral hydrogen. The reduction in
the average number of scattering events can, for example, lead to
a decrease in the effectiveness of the Wouthuysen--Field coupling \citep{Wouthuysen1952,Field1958,Field1959}
of the spin temperature to Ly$\alpha$ radiation or affect the polarization
of the scattered photons, since both are influenced by the number
of resonant scattering events \citep{Roy2009,Dijkstra2008,SangHteon2015,Seon2020,Seon2022}.

We conclude that modeling turbulence as an effective temperature (microturbulence)
has a limited area of applicability. On one hand, our study confirms
the importance of coupling radiative transfer codes with detailed
hydrodynamic simulations \citep{Smith2020}. On the other hand, our
approach provides a simplified alternative when the use of full hydrodynamic
simulations is limited by computational resources or by the availability
of reliable physical inputs. Thus, our model is especially useful
in the regions where the scale of the turbulence is too large to be
described as microturbulence but at the same time is too small to
be realistically resolved by full hydrodynamic simulations.

Finally, we point out that the propagation of Ly$\alpha$ photons
can have an effect on the macroscopic motion of the neutral gas itself.
Indeed, Ly$\alpha$ photons can transfer momentum between layers of
the moving gas equilibrating their relative motion through radiative
viscosity similar to the case of radiative viscosity due to Thomson
scattering~\citep{Loeb1992}, with the crucial difference being that
unlike Thomson scattering, the mean free path of the Ly$\alpha$ photons
is not constant but sensitively, by orders of magnitude, depends on
the photon wavelength.

\section*{Acknowledgements}

Most of the work has been done while VM was at Princeton University
and AK at IAS. We acknowledge the cluster resources provided by IAS
for computer simulations performed in this paper.

%%%%%%%%%%%%%%%%%%%%%%%%%%%%%%%%%%%%%%%%%%%%%%%%%%
\section*{Data Availability}
The code and some data underlying this article are available in the GitHub repository \citet{Munirov_LyAMC_2022} at \url{https://github.com/dimmun/LyAMC}.
The additional data underlying this article will be shared on reasonable request to the corresponding author.

%%%%%%%%%%%%%%%%% APPENDICES %%%%%%%%%%%%%%%%%%%%%

\appendix

\onecolumn

%%%%%%%%%%%%%%%%%%%%%%%%%%%%%%%%%%%%%%%%%%%%%%%%%%%
\newpage{}
\section{Acceptance rate for get\_upar}
To pick a random parallel velocity for scattering atom we use the
rejection method. In this method we use parameter $u_{0}$ that determines
acceptance rate and thus can accelerate computations if chosen wisely.
Here is the comparison for the acceptance rate for the choice of $u_{0}$
used in our paper with \citep{Semelin2007} and \citep{Laursen2009,Laursen2010}
\begin{figure*}
\includegraphics[width=0.7\textwidth]{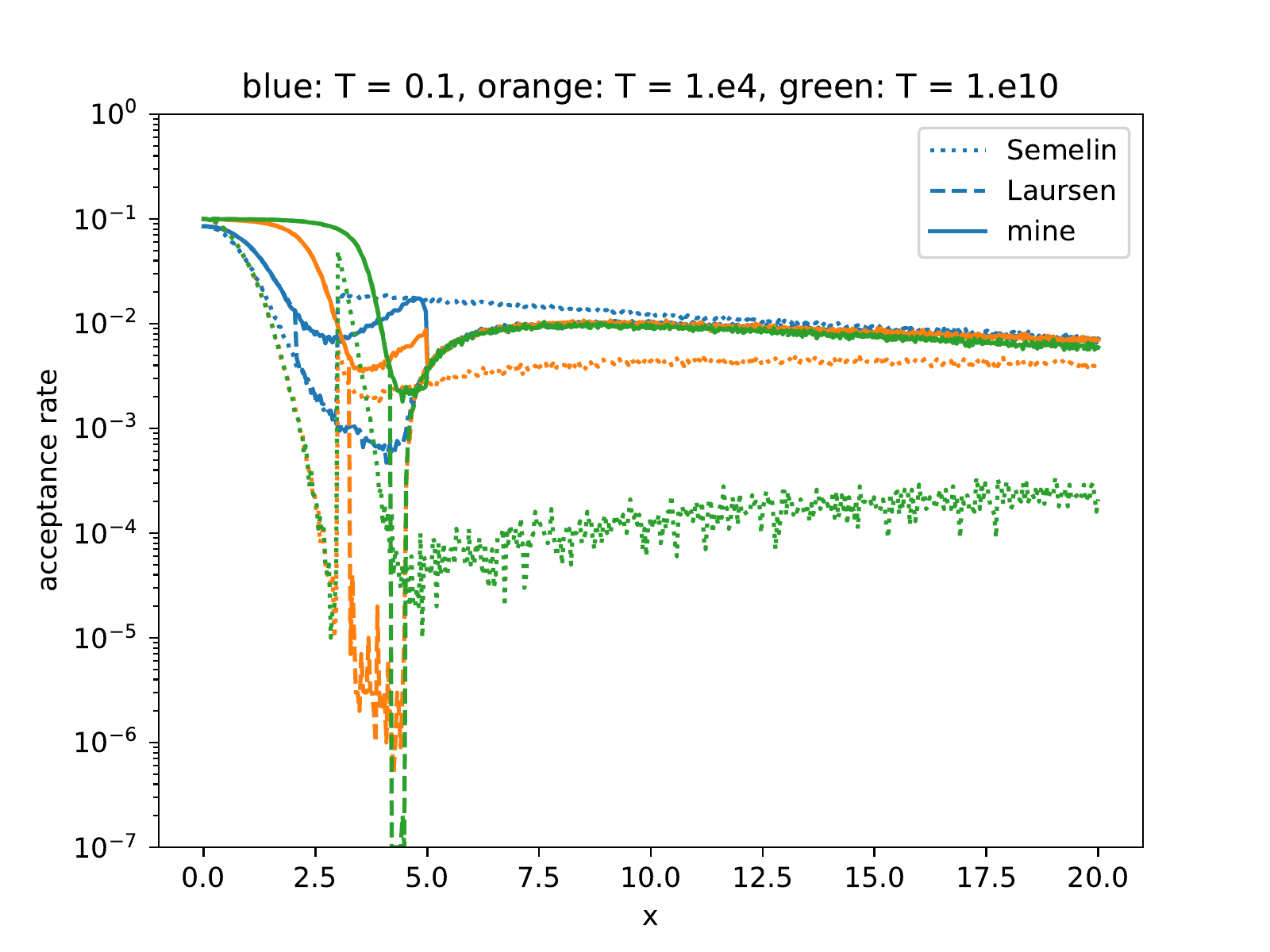}
\caption{The acceptance rate for the choice of $u_{0}$
used in our paper with \citep{Semelin2007} and \citep{Laursen2009,Laursen2010}
for three different values of $T$ and fixed $\tau_{0}$ (consequently
for three different values of parameter $a\tau_{0}$).}
\end{figure*}

%%%%%%%%%%%%%%%%%%%%%%%%%%%%%%%%%%%%%%%%%%%%%%%%%%%
\newpage{}
\section{Static cloud}
The numerical solution for the case of static uniform cloud with a
source in the center of a uniform sphere of static gas cloud emitting
photons at line center with analytical results.
\begin{figure*}
\includegraphics[width=0.7\textwidth]{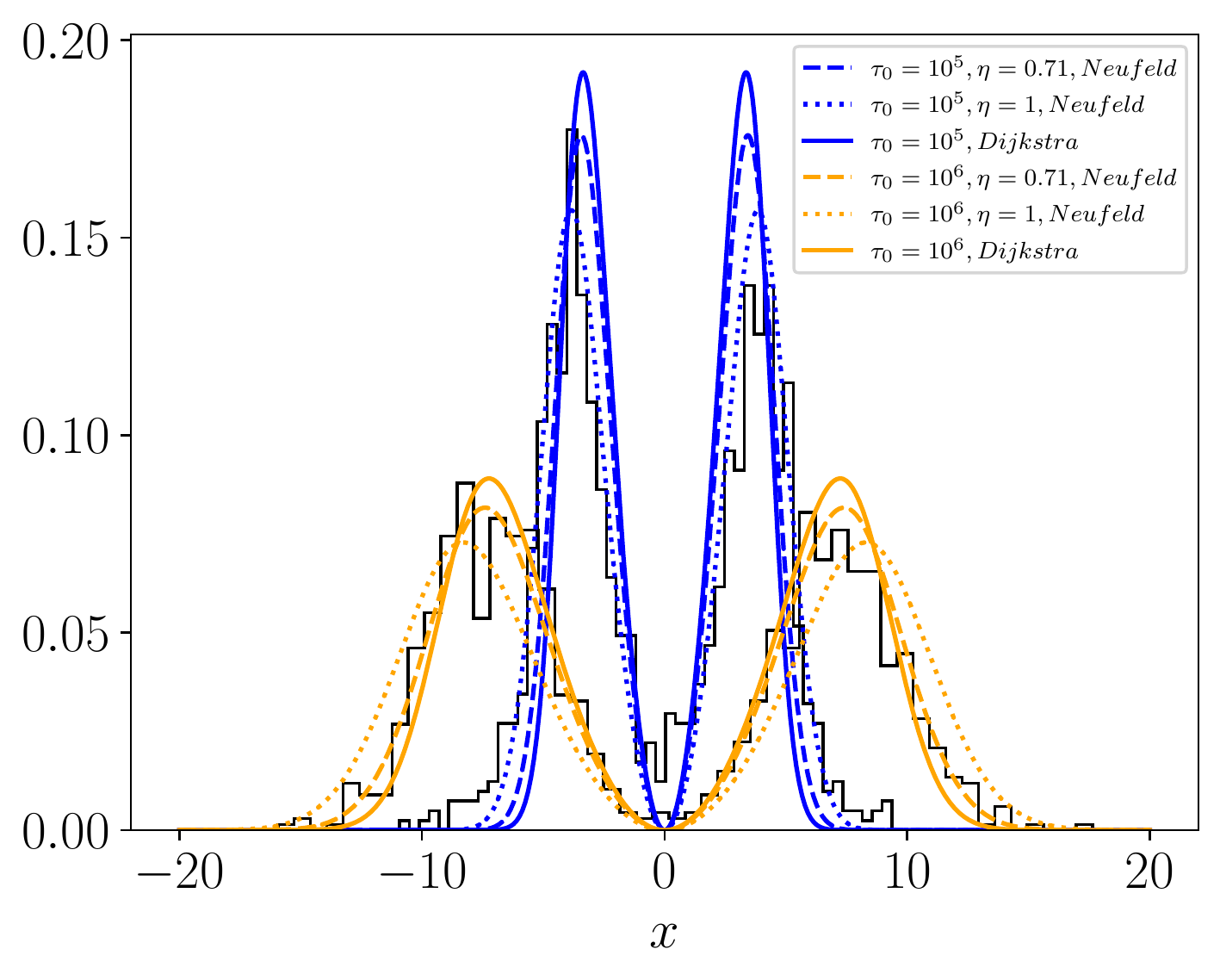}
\caption{The emergent spectrum from a static uniform sphere
for temperature $T=10^{4}\:\textrm{K}$ and two optical depths $\tau_{0}=10^{5}$
and $\tau_{0}=10^{6}$. The figure also shows the analytical solutions
from Neufeld \citep{Neufeld1990} using $\eta a\tau_{0}$ with $\eta=0.71$
(dashed line) and $\eta=1$ (dotted line) and from Dijkstra \citep{Dijkstra2006a}
(solid line).}
\end{figure*}

%%%%%%%%%%%%%%%%%%%%%%%%%%%%%%%%%%%%%%%%%%%%%%%%%%
\newpage{}
\section{Expanding/contracting clouds}
The numerical solution for the case of expanding and contracting uniform
cloud with a source in the center of a uniform sphere of neutral hydrogen
cloud emitting photons at line center. See Ref.~\citet{Zheng2002}.
\begin{figure*}[H]
\includegraphics[width=0.7\textwidth]{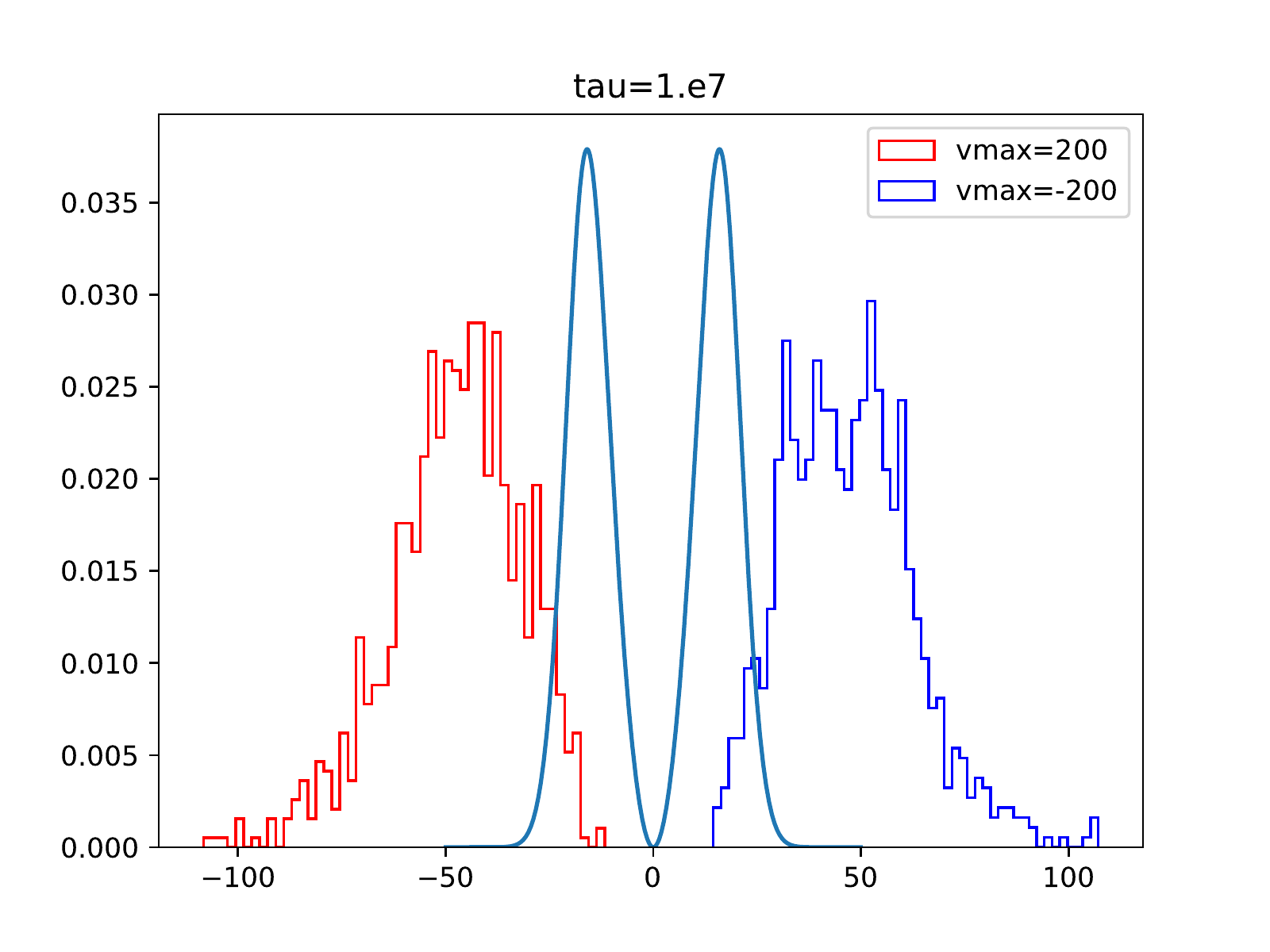}
\caption{The emergent spectrum from expanding
($v=200\:\textrm{km}/\textrm{s}$) and contracting ($v=-200\:\textrm{km}/\textrm{s}$)
uniform spheres for optical depth $\tau_{0}=10^{7}$.}
\end{figure*}

%%%%%%%%%%%%%%%%%%%%%%%%%%%%%%%%%%%%%%%%%%%%%%%%%%
\newpage{}
\section{Density gradient model}
The density gradient model of \citet{Zheng2014}. See Figs.~1 and
3 from \citet{Zheng2014}.
\begin{figure*}[H]
\includegraphics[width=0.5\textwidth]{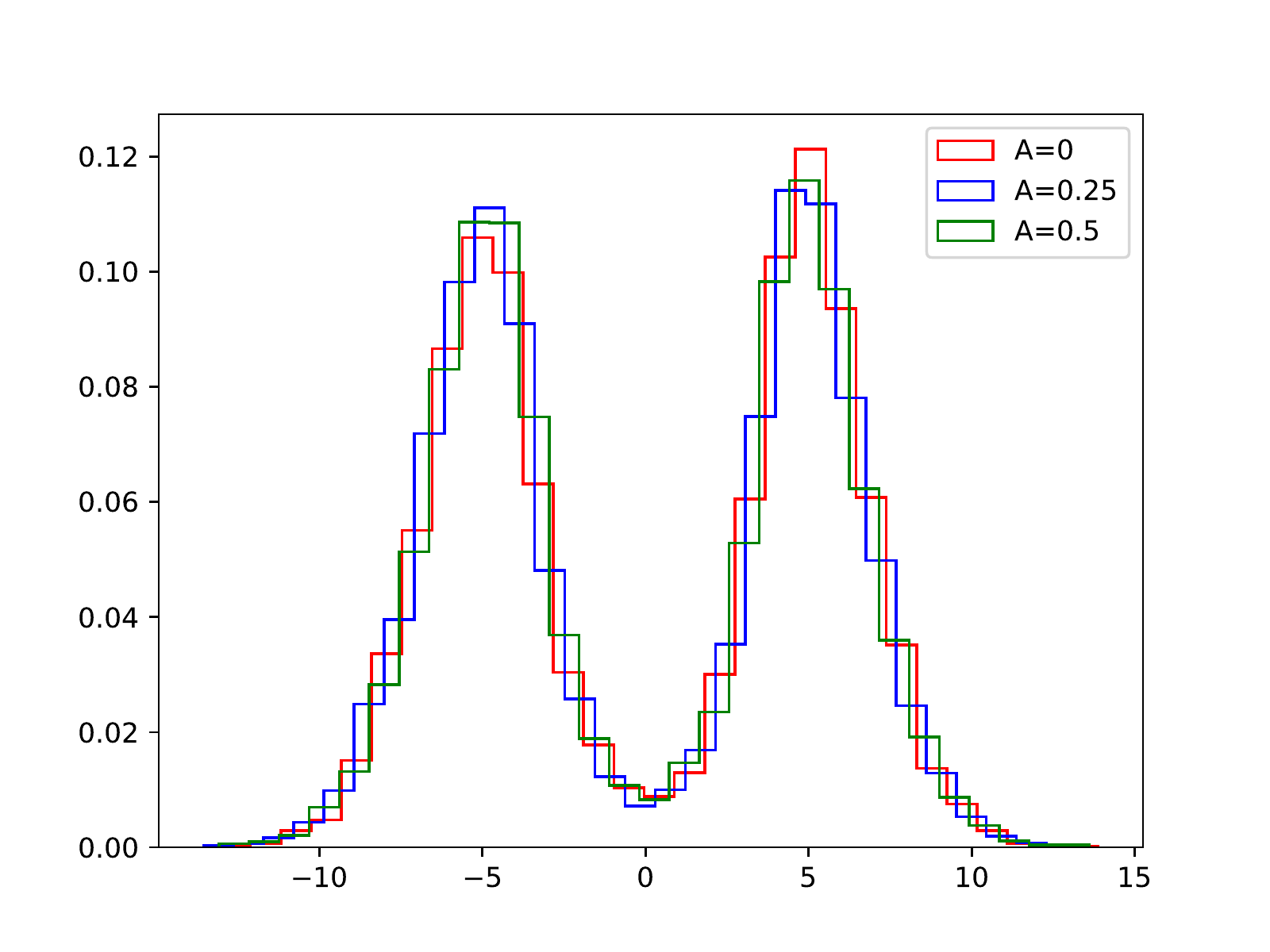}\includegraphics[width=0.5\textwidth]{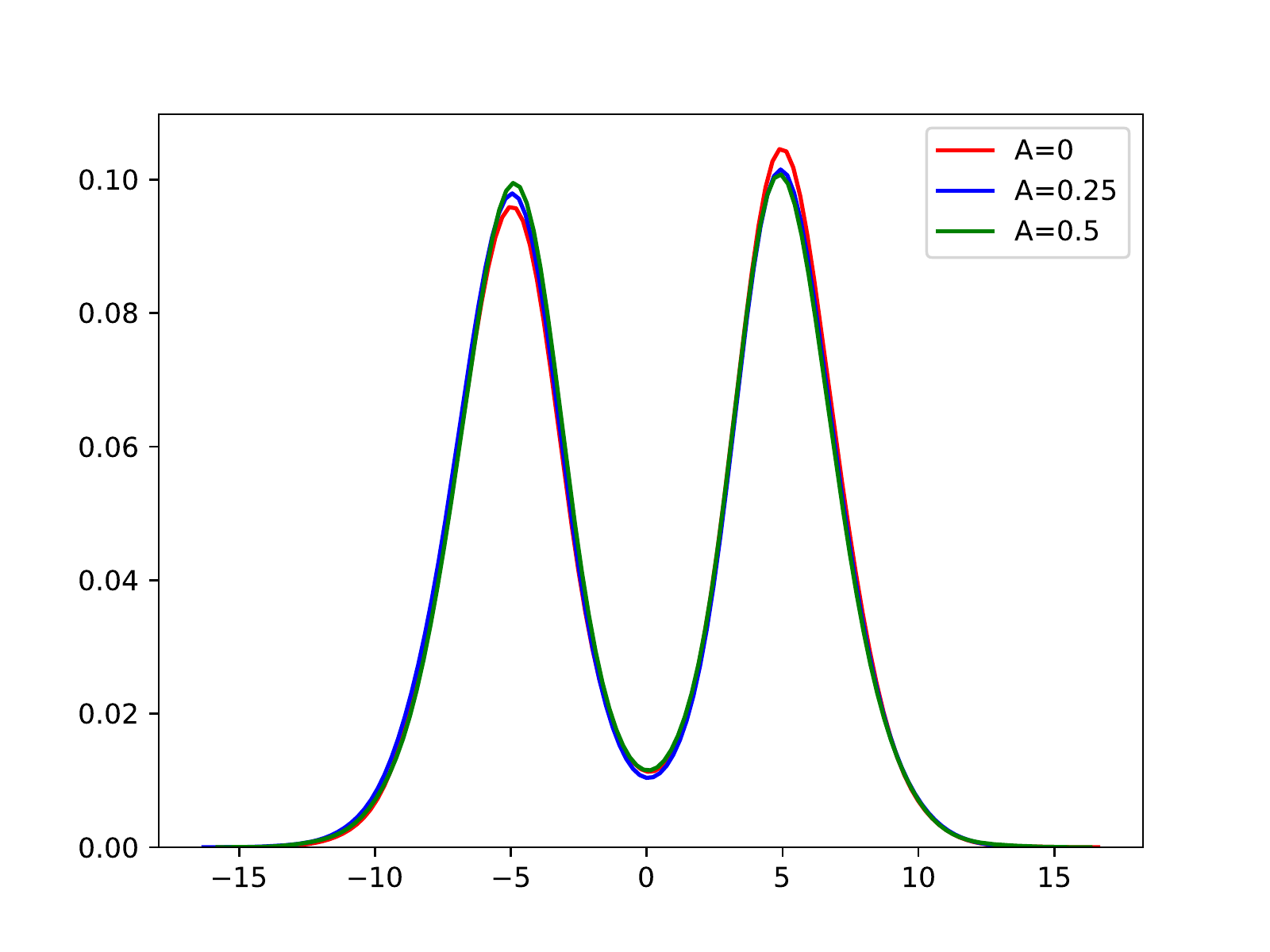}
\includegraphics[width=0.5\textwidth]{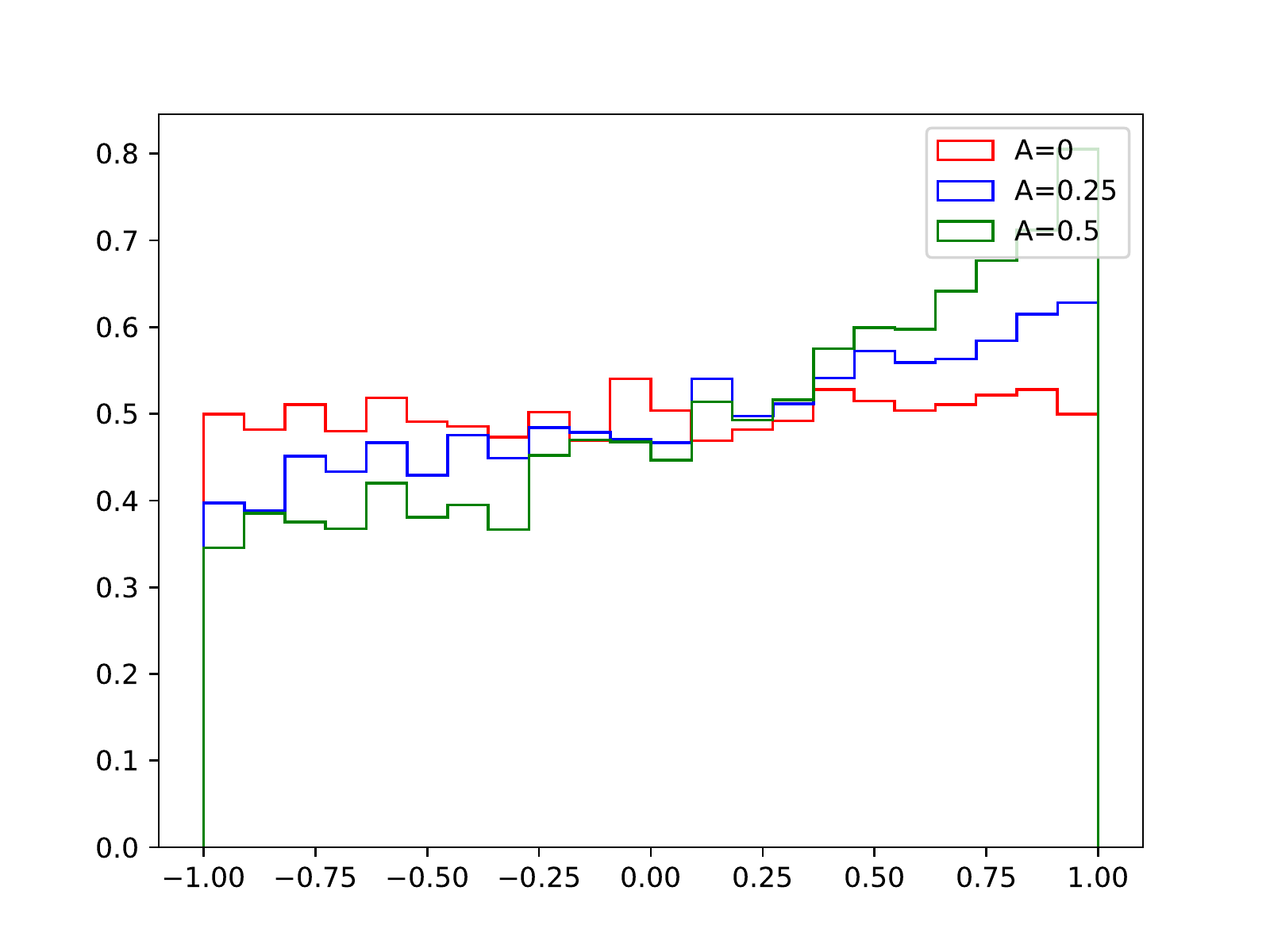}\includegraphics[width=0.5\textwidth]{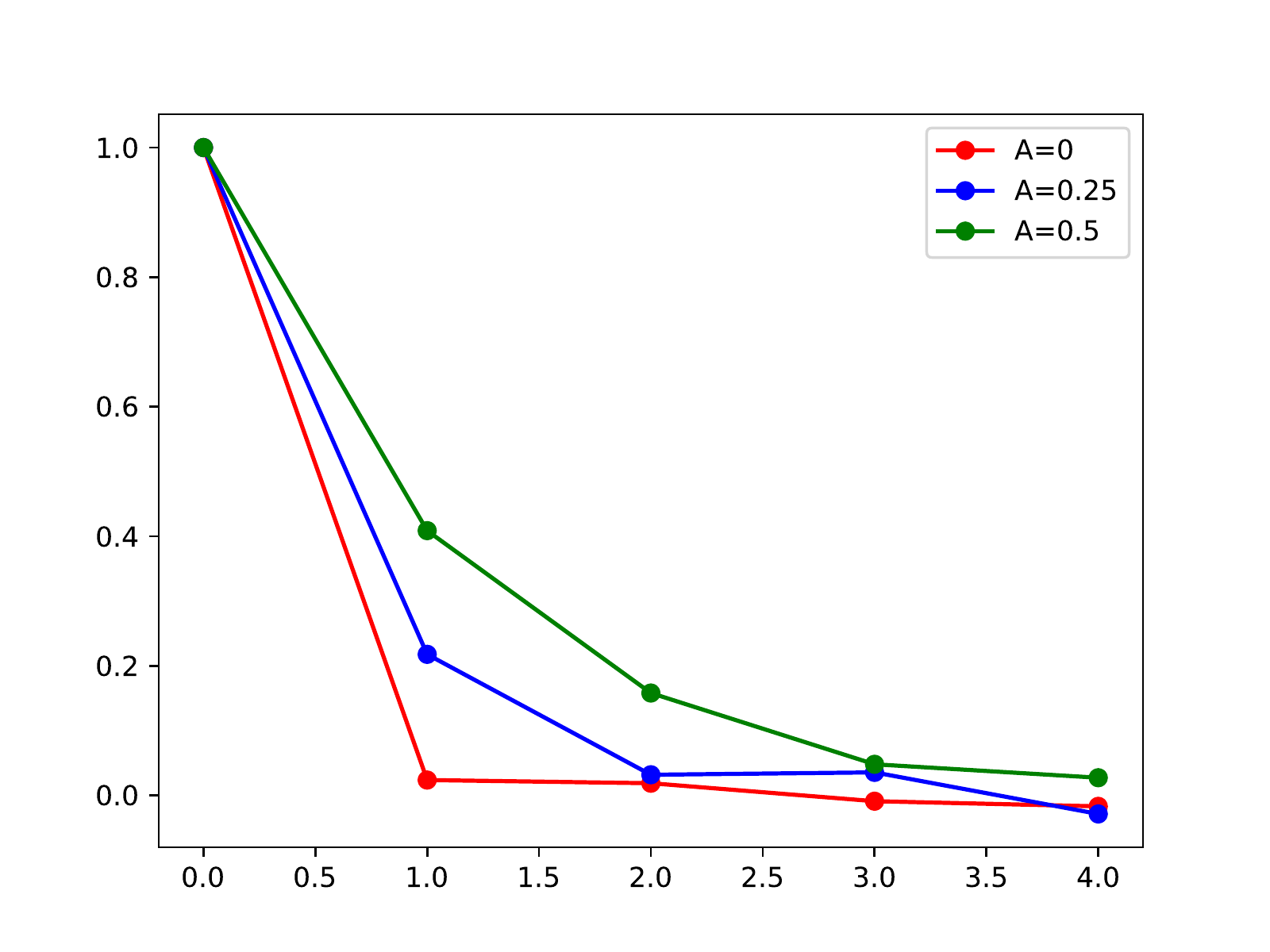}
\caption{Results for the density gradient model (see \citealt{Zheng2014}).}
\end{figure*}

%%%%%%%%%%%%%%%%%%%%%%%%%%%%%%%%%%%%%%%%%%%%%%%%%%
\newpage{}
\section{Velocity gradient model}
The velocity gradient model of~\citet{Zheng2014}. See Figs.~5 and
7 from \citet{Zheng2014}.
There is some disagreement but the corresponding graph in \citet{Zheng2014}
was actually computed for the parameters different from those reported
in~\citet{Zheng2014} {[}private communication{]}. 
\begin{figure*}[H]
\includegraphics[width=0.5\textwidth]{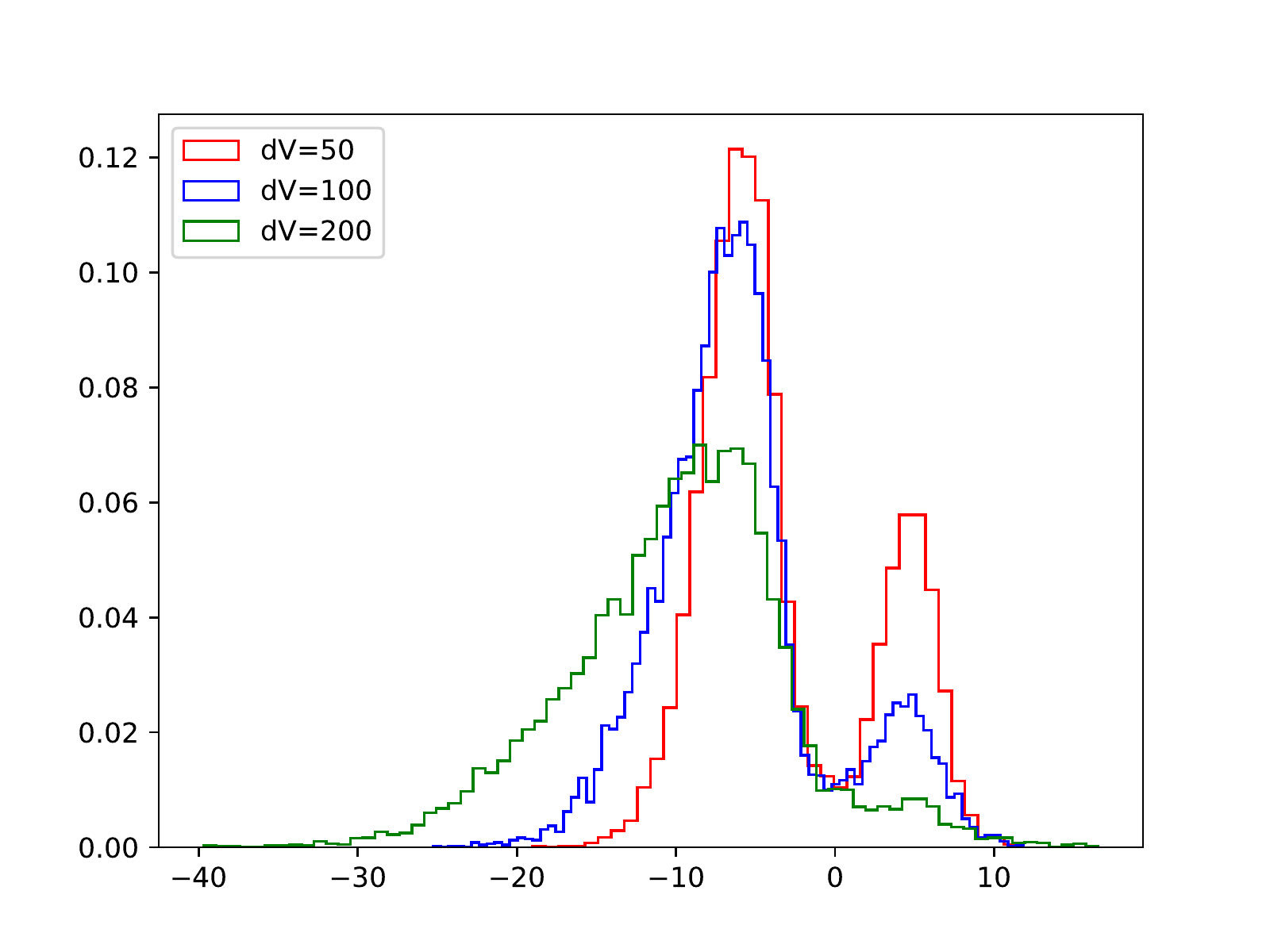}\includegraphics[width=0.5\textwidth]{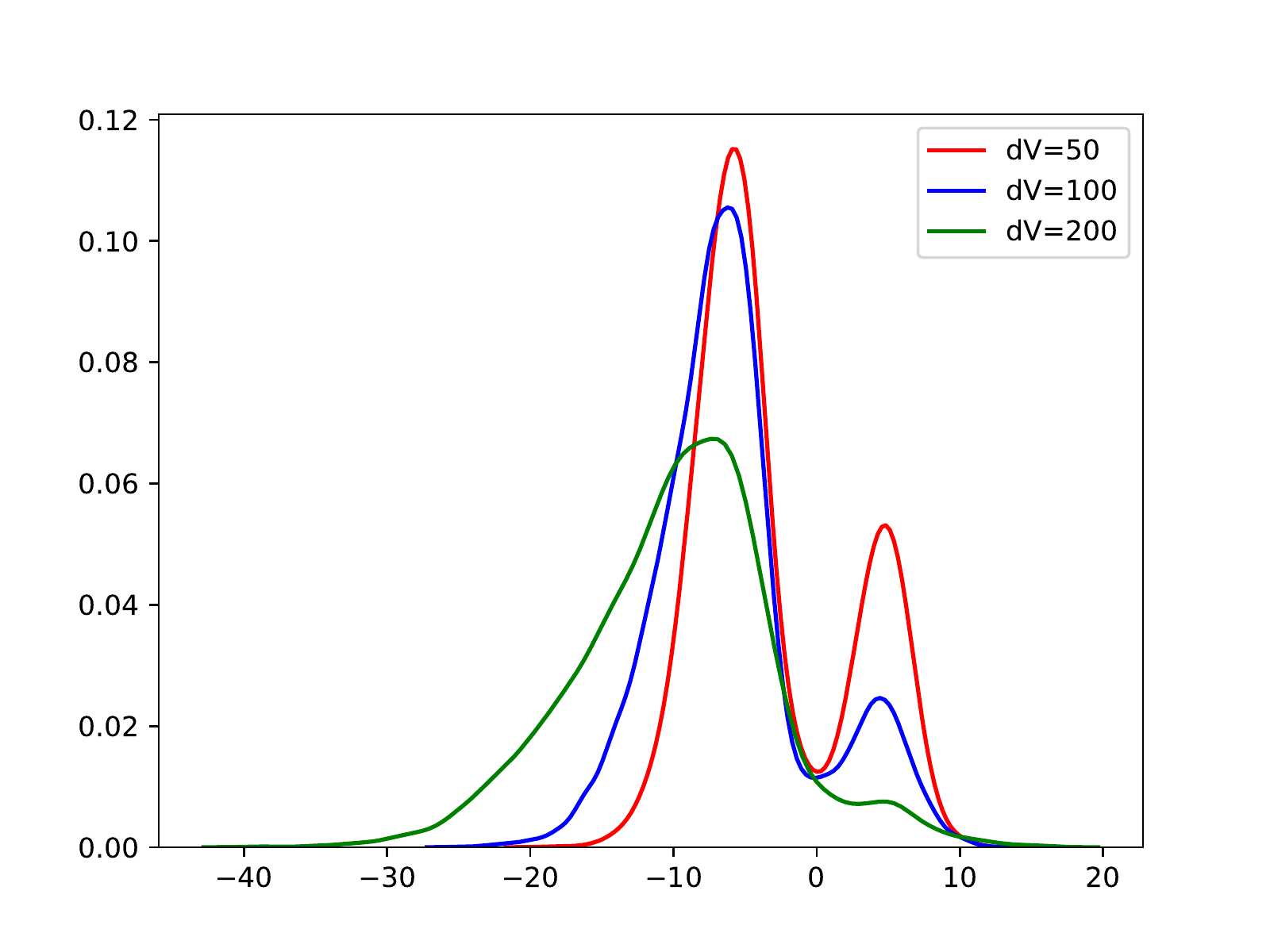}
\includegraphics[width=0.5\textwidth]{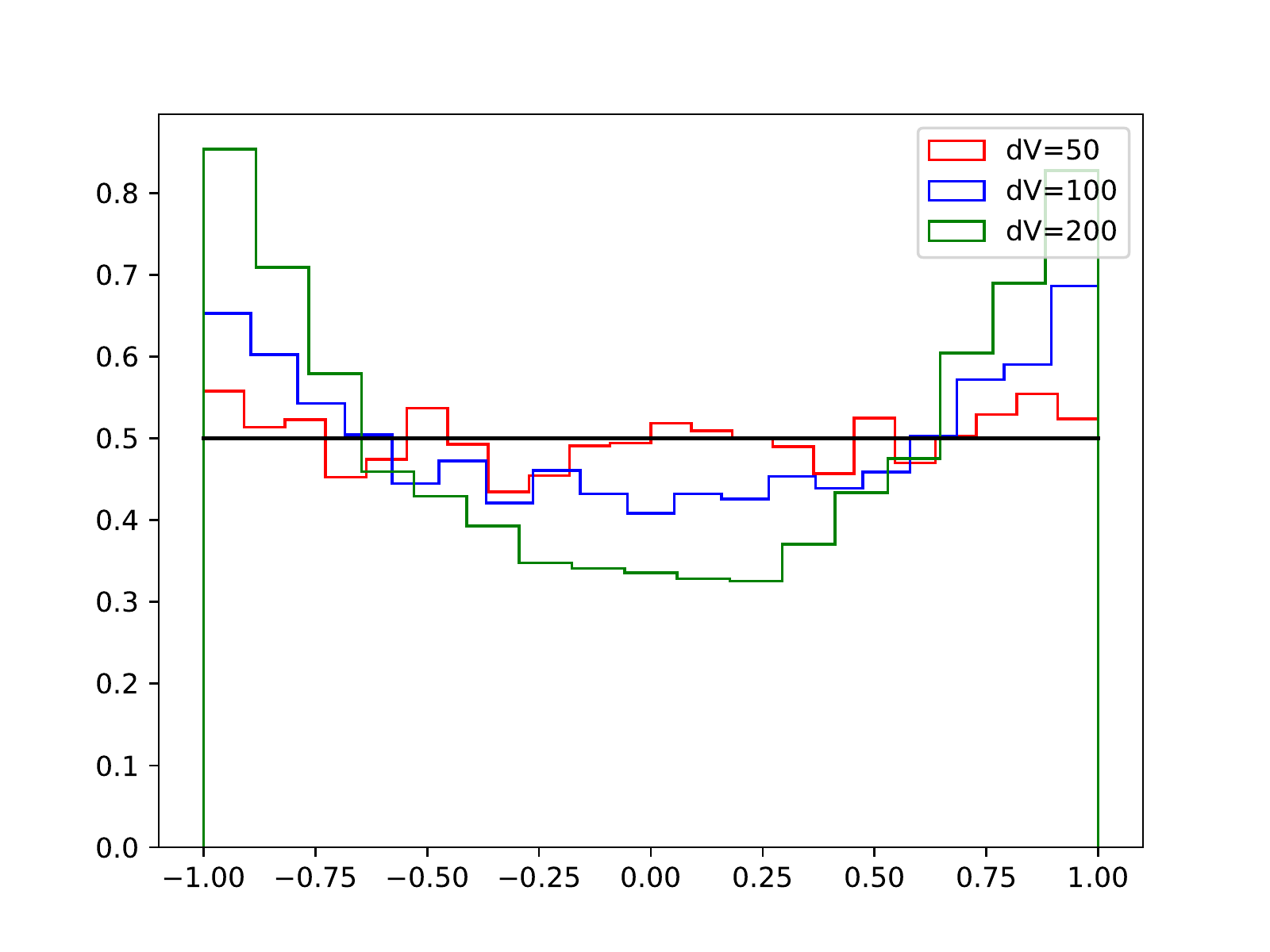}\includegraphics[width=0.5\textwidth]{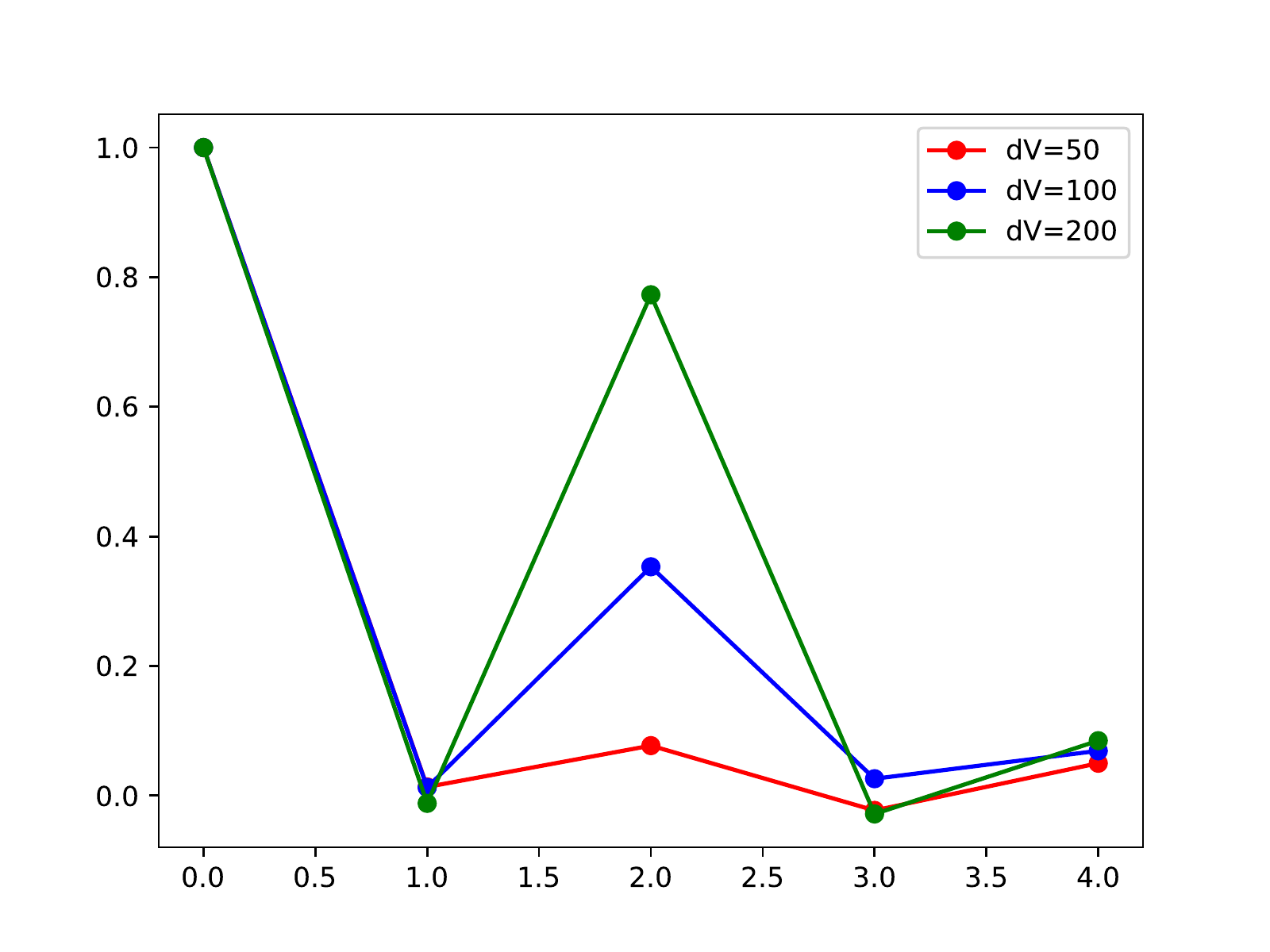}
\caption{Results for the velocity gradient model (see \citealt{Zheng2014}).}
\end{figure*}

%%%%%%%%%%%%%%%%%%%%%%%%%%%%%%%%%%%%%%%%%%%%%%%%%%
\newpage{}
\section{Bipolar wind model}
The bipolar wind model of~\citet{Zheng2014}. See Fig.~9 from~\citet{Zheng2014}.
\begin{figure*}[H]
\includegraphics[width=0.5\textwidth]{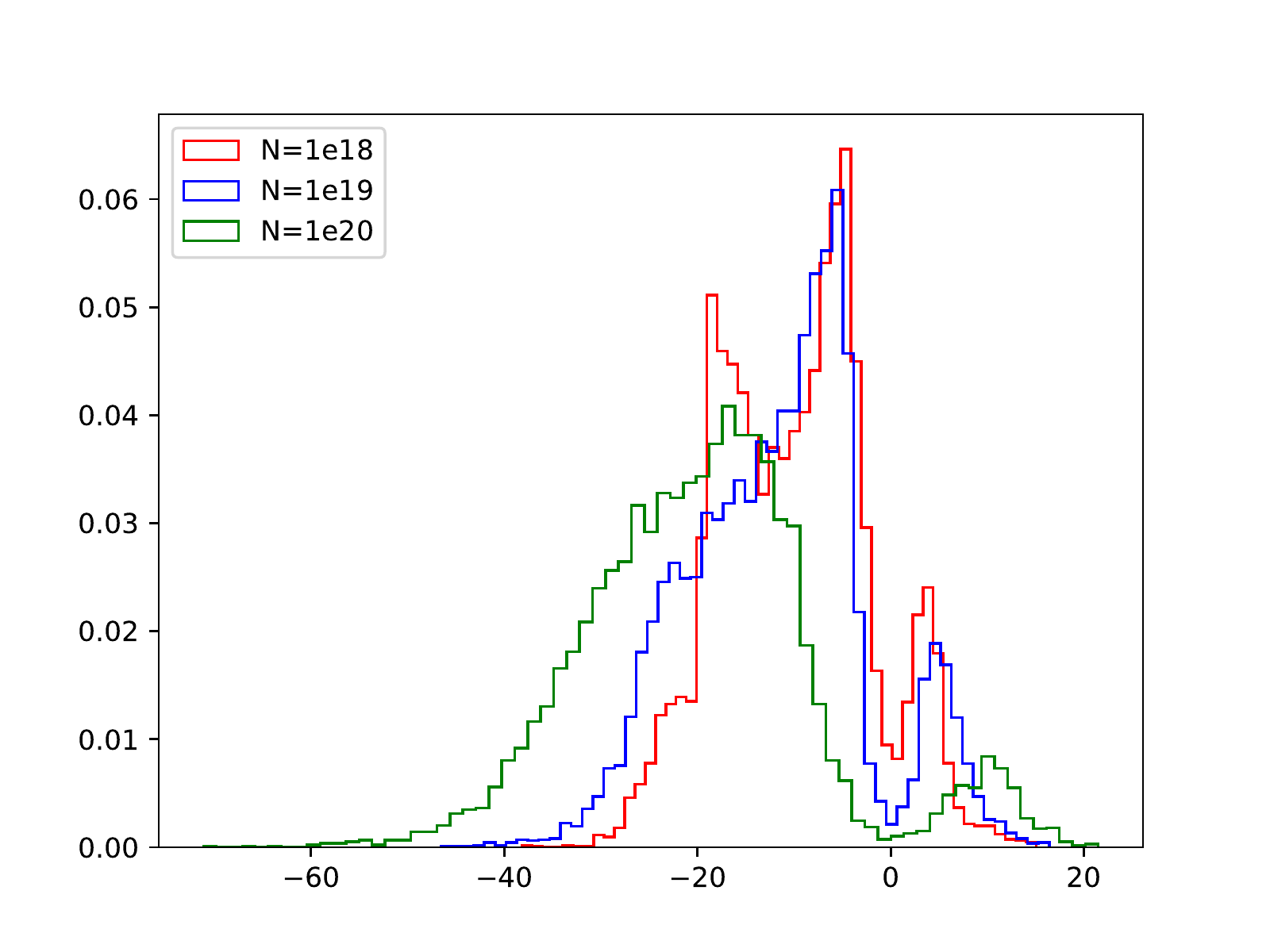}\includegraphics[width=0.5\textwidth]{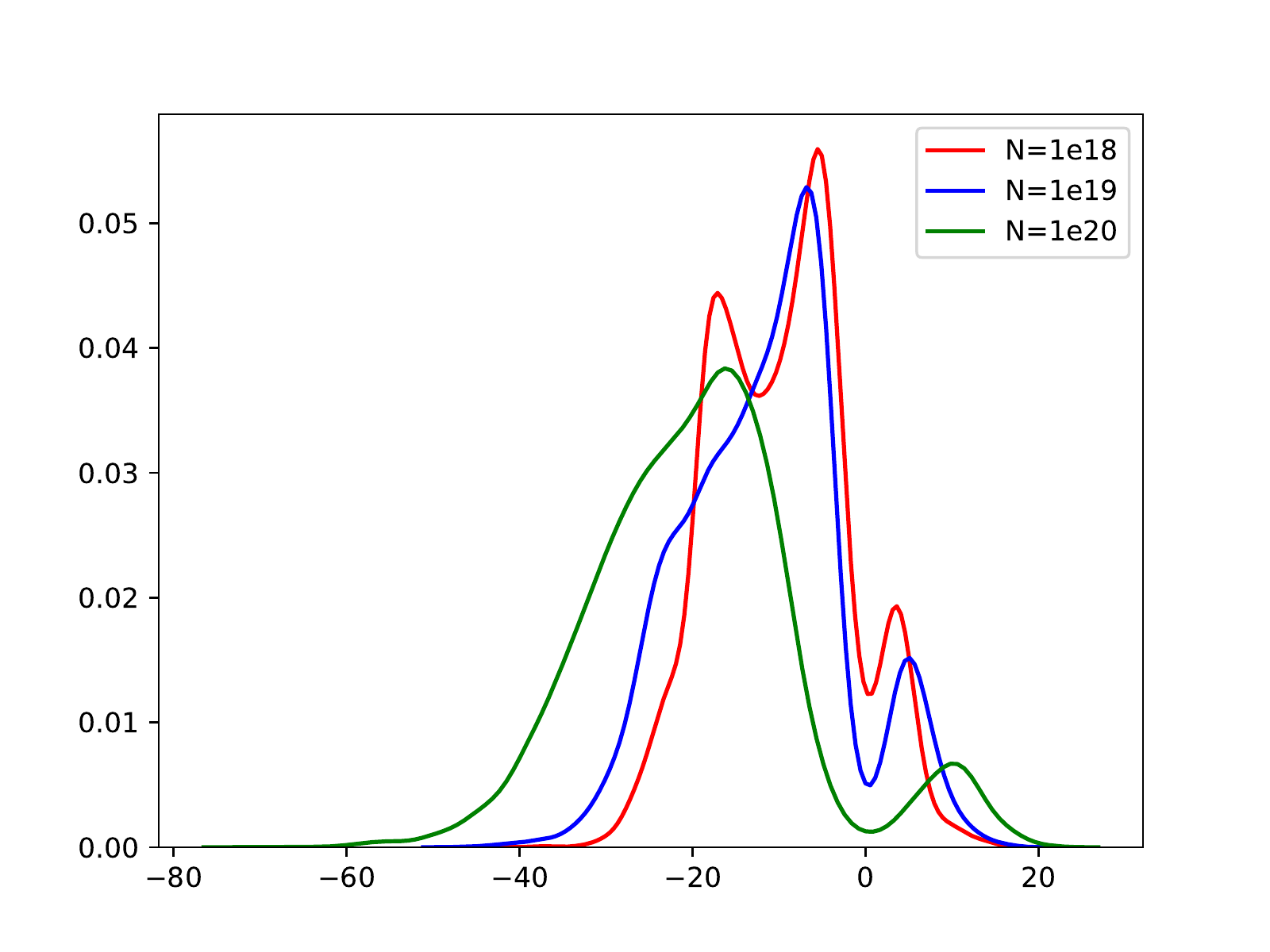}
\includegraphics[width=0.5\textwidth]{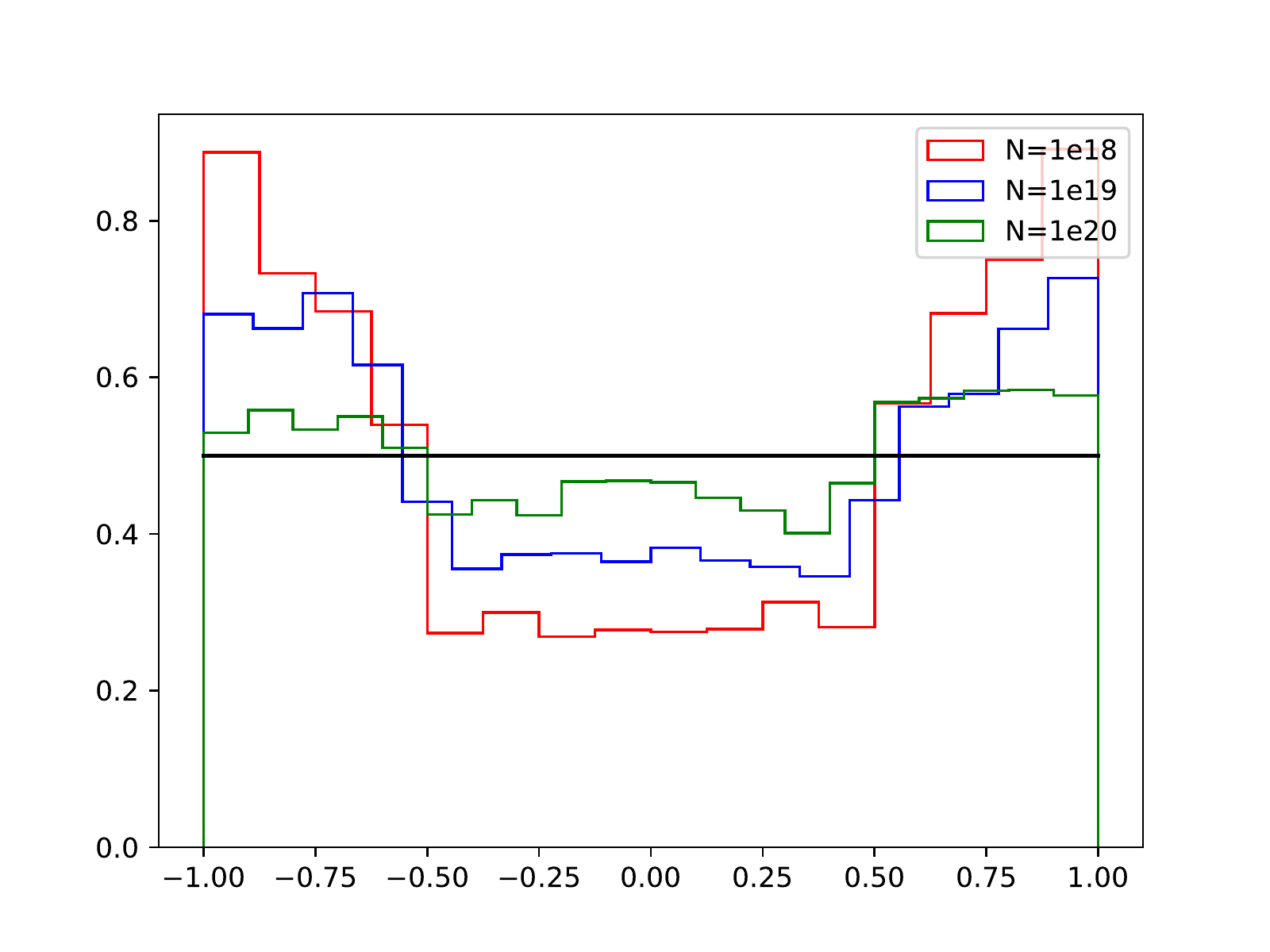}\includegraphics[width=0.5\textwidth]{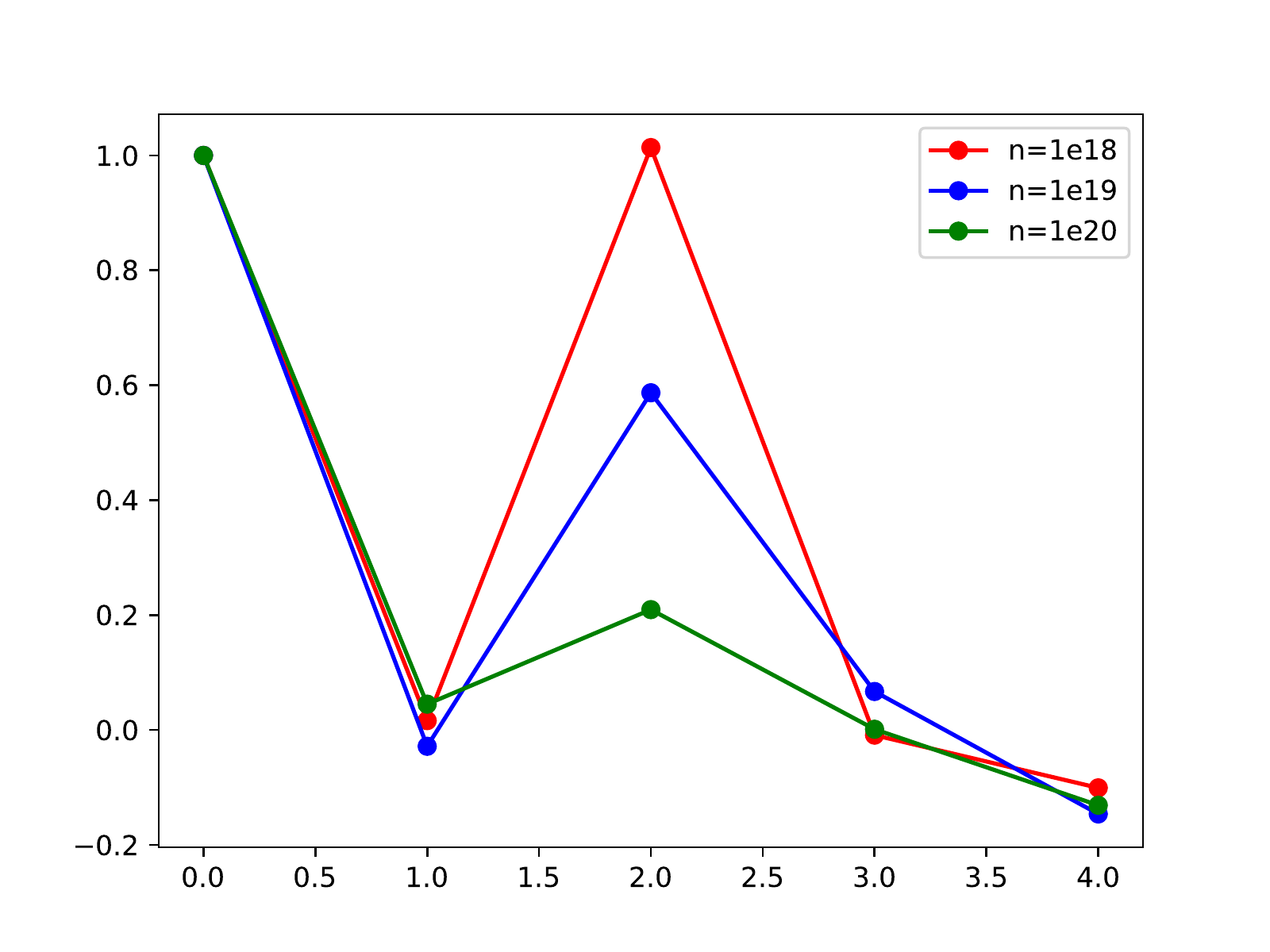}
\caption{Results for the bipolar wind model (see \citealt{Zheng2014}).}
\end{figure*}

%%%%%%%%%%%%%%%%%%%%%%%%%%%%%%%%%%%%%%%%%%%%%%%%%%
\newpage{}
\section{The number of scatterings $N_{\textrm{scat}}$ for $\tau_{0}=10^{5}$,
$T=1\:\textrm{K}$, $a\tau_{0}\approx4.7\times10^{3}$}
The number of scatterings $N_{\textrm{scat}}$ for $\tau_{0}=10^{5}$,
$T=1\:\textrm{K}$, $a\tau_{0}\approx4.7\times10^{3}$.
\begin{figure*}[H]
\includegraphics[width=0.5\textwidth]{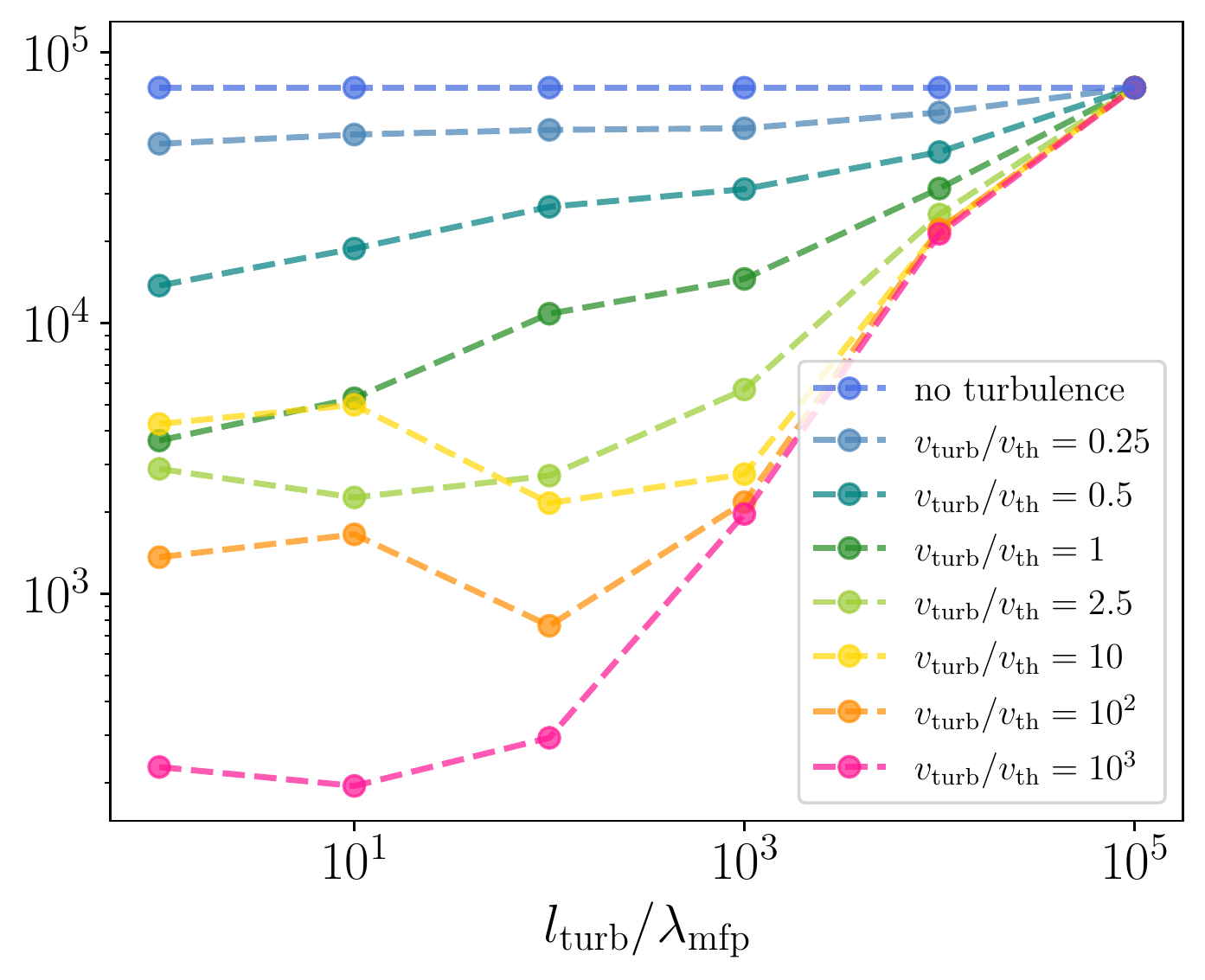}\includegraphics[width=0.5\textwidth]{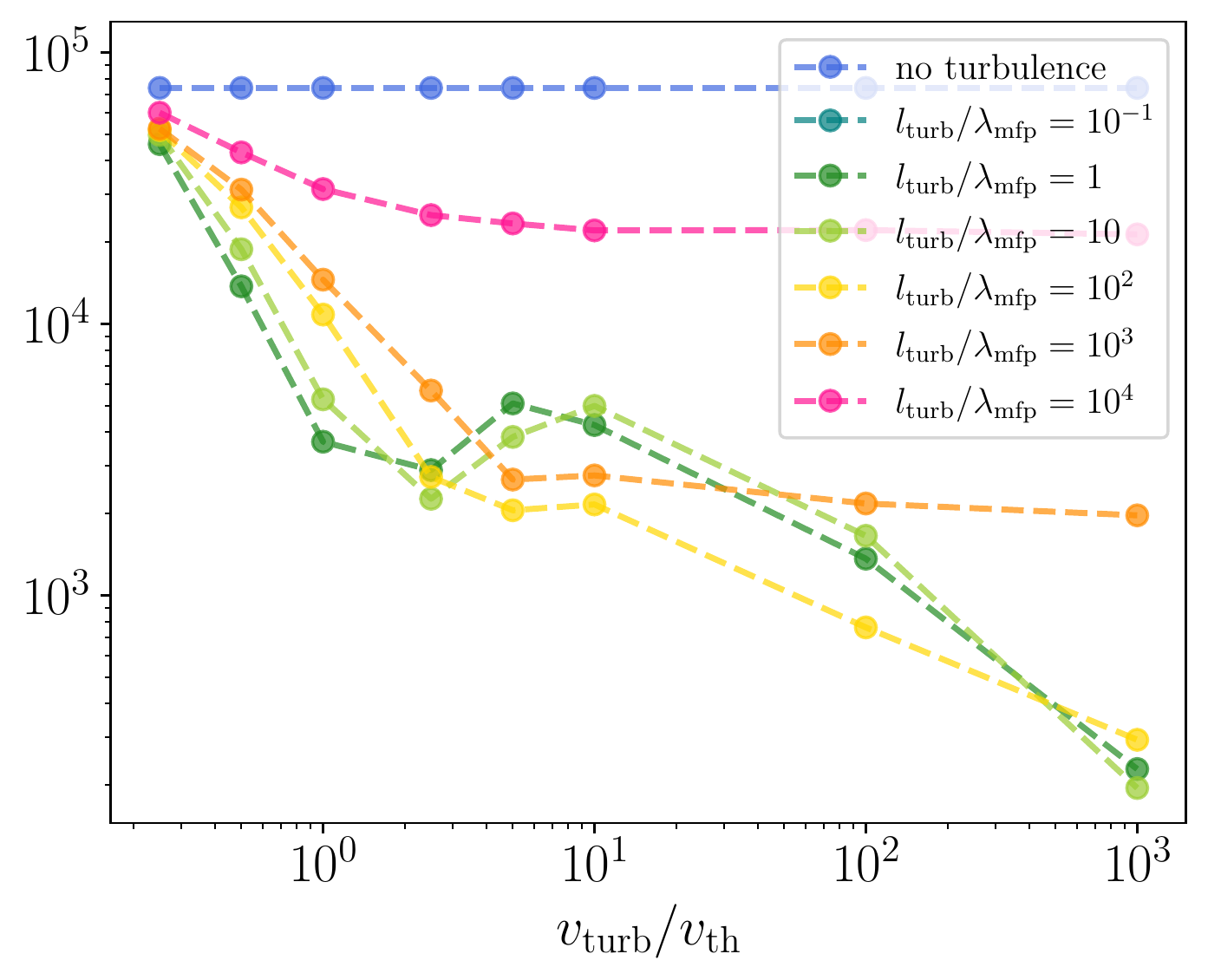}
\caption{Left panel:
The average number of scattering events $N_{\textrm{scat}}$ that
a representative photon undergoes before it escapes the turbulent
cloud of neutral hydrogen versus the turbulence correlation length
$l_{\textrm{turb}}$ for different values of the turbulence velocity
amplitude $v_{\textrm{turb}}$. Right panel: The average number of
scattering events $N_{\textrm{scat}}$ that a representative photon
undergoes versus the turbulence velocity $v_{\textrm{turb}}$ for
different values of the turbulence correlation length $l_{\textrm{turb}}$.
The temperature is $T=1\:\textrm{K}$ and the optical depth is $\tau_{0}=10^{5}$,
which corresponds to a strongly optically thick case with the Voigt
parameter $a$ such that $a\tau_{0}\approx4.7\times10^{3}$.}
\end{figure*}

%%%%%%%%%%%%%%%%%%%%%%%%%%%%%%%%%%%%%%%%%%%%%%%%%%
\newpage{}
\section{The effective mean free path for $\tau_{0}=10^{5}$, $T=1\:\textrm{K}$,
$a\tau_{0}\approx4.7\times10^{3}$}
The effective mean free path for $\tau_{0}=10^{5}$, $T=1\:\textrm{K}$,
$a\tau_{0}\approx4.7\times10^{3}$.
\begin{figure*}[B]
\includegraphics[width=0.3\paperwidth]{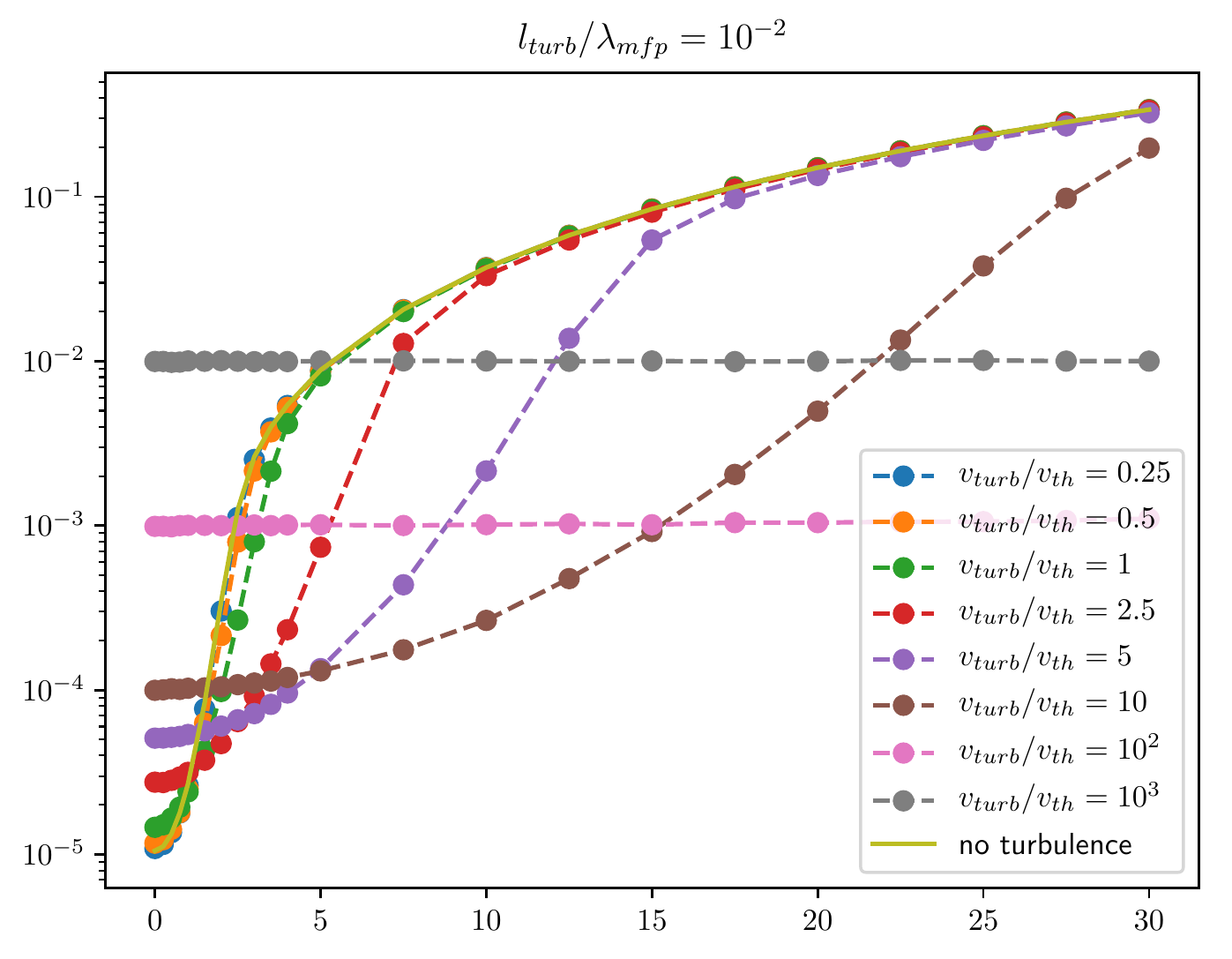}\includegraphics[width=0.3\paperwidth]{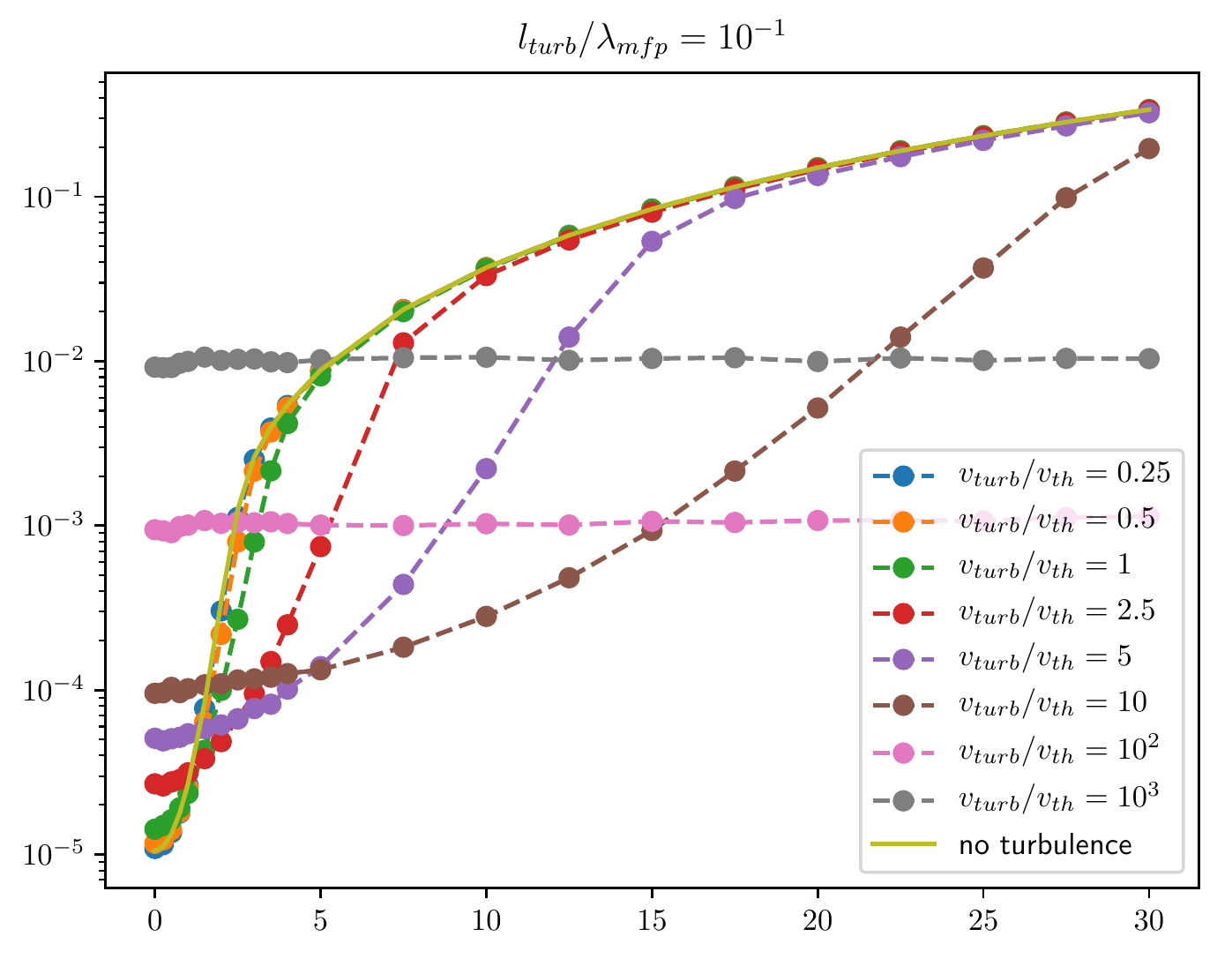}\includegraphics[width=0.3\paperwidth]{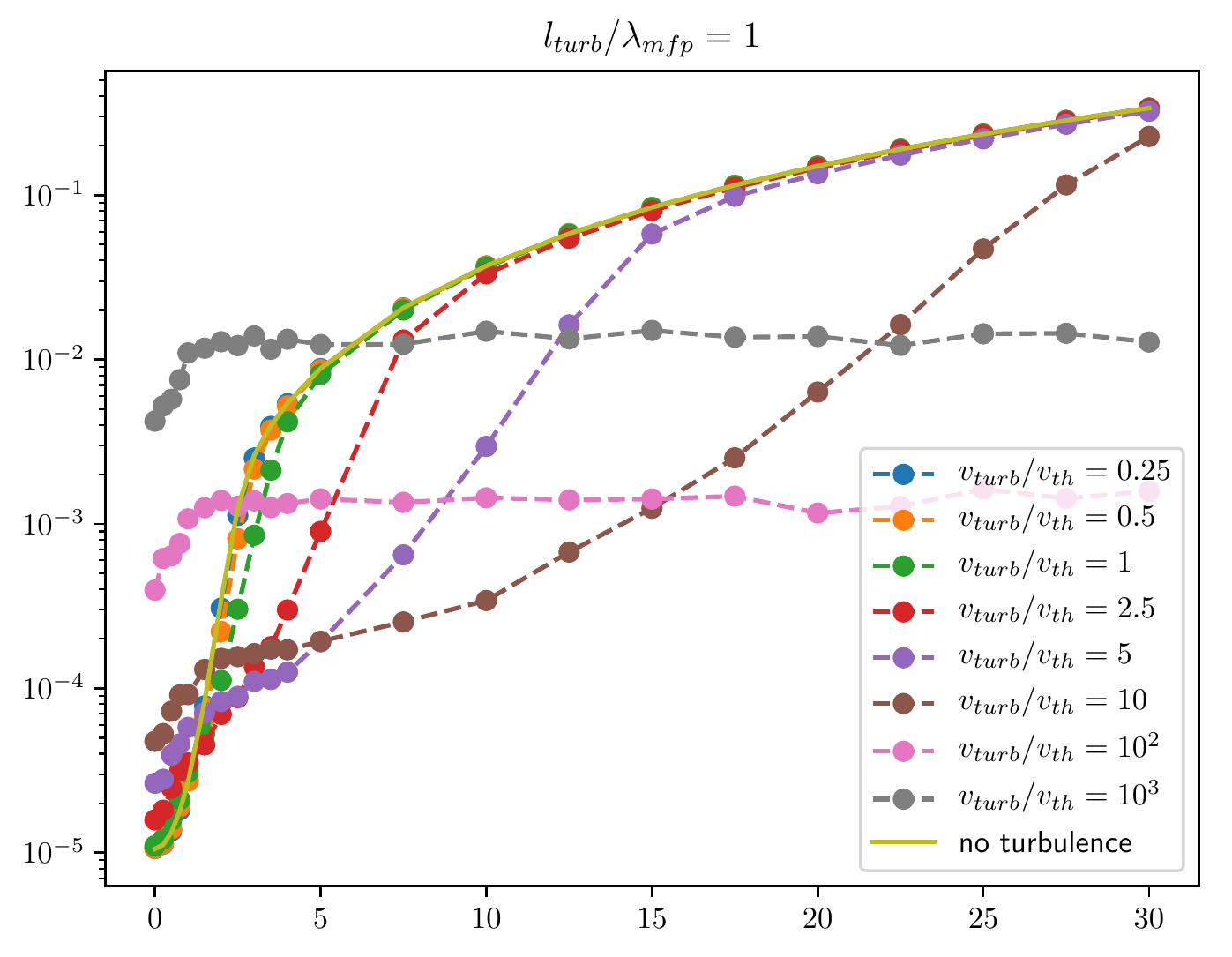}
\includegraphics[width=0.3\paperwidth]{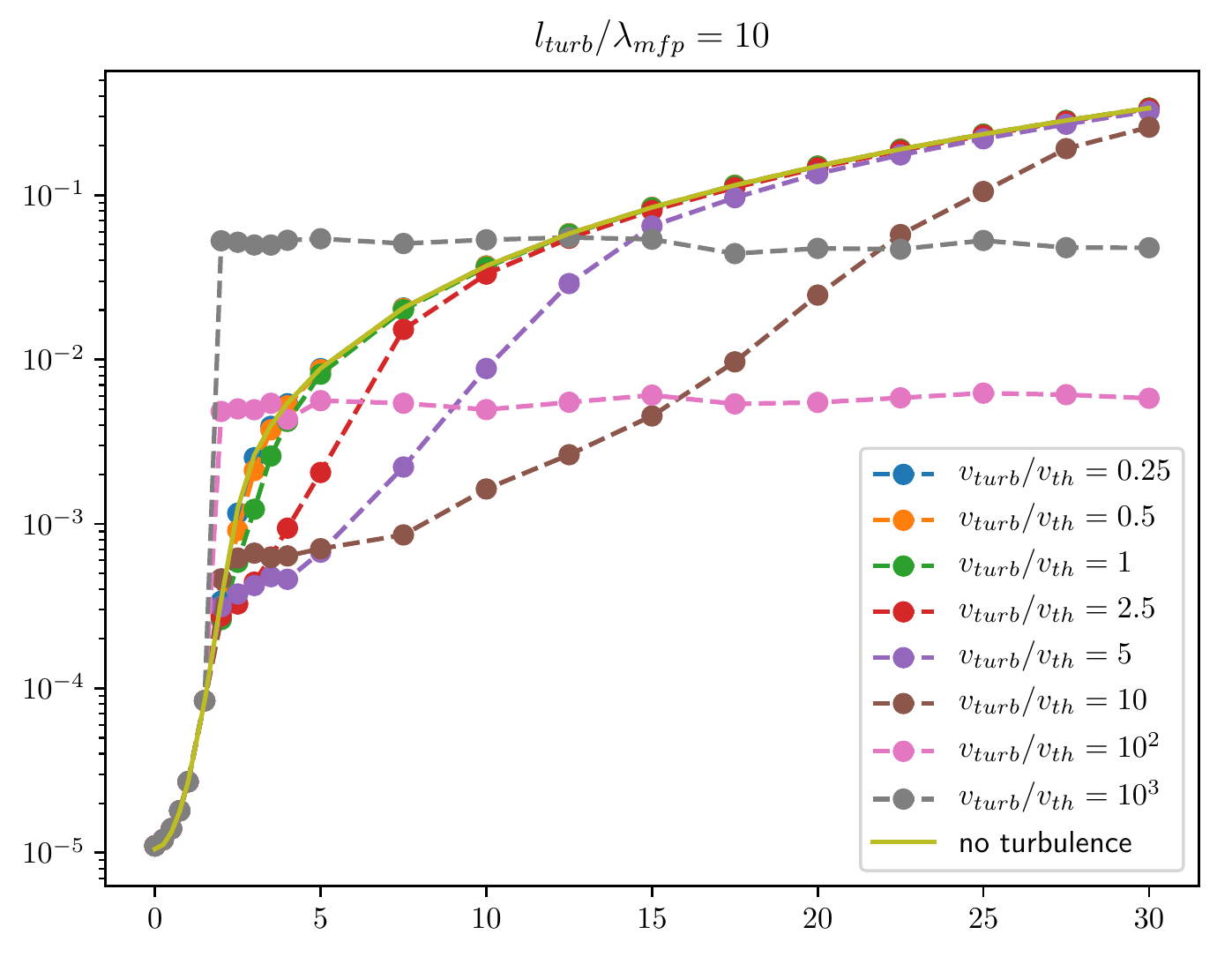}\includegraphics[width=0.3\paperwidth]{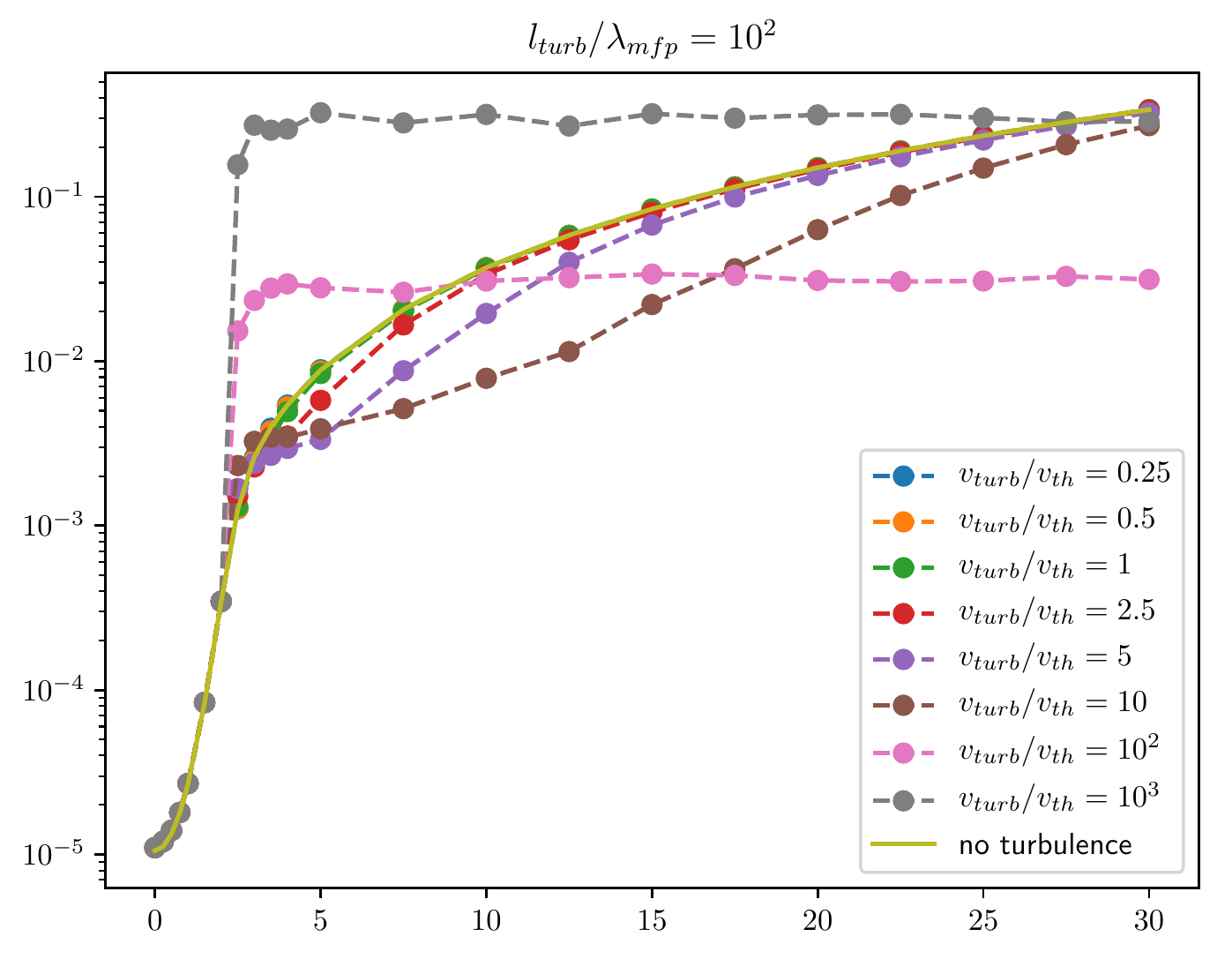}\includegraphics[width=0.3\paperwidth]{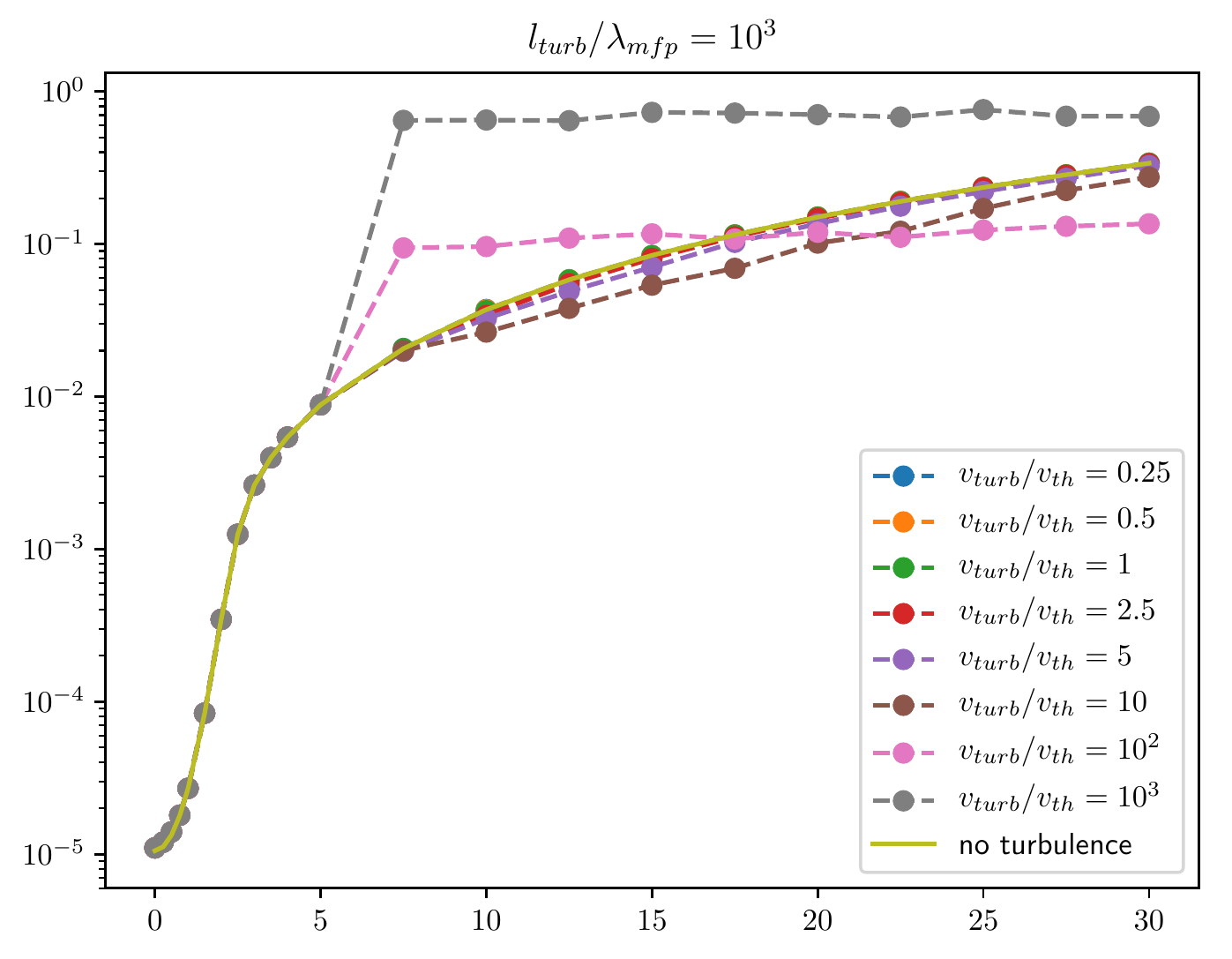}
\includegraphics[width=0.3\paperwidth]{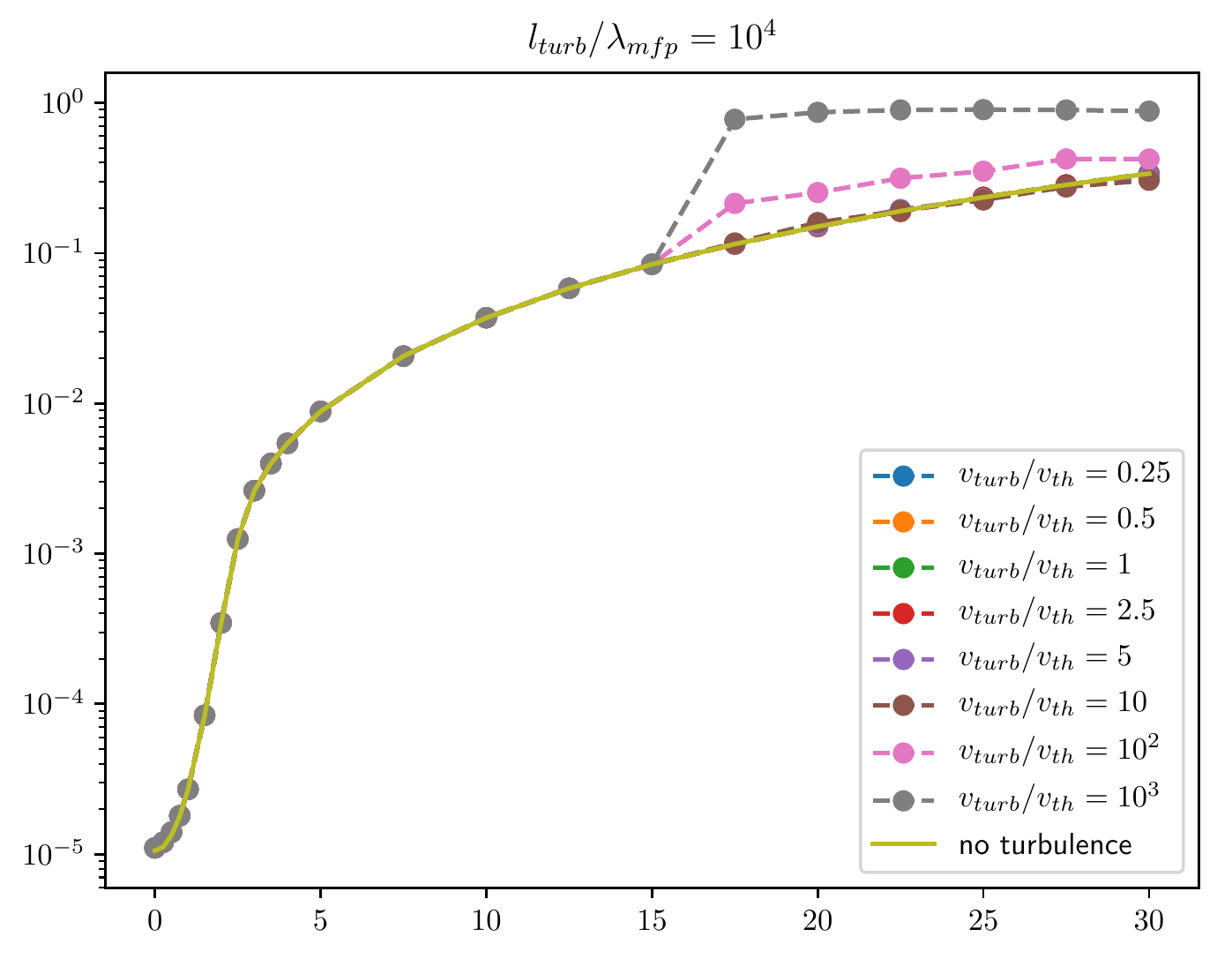}
\caption{The numerically calculated effective mean free path versus the dimensionless
frequency $x$ for $\tau_{0}=10^{5}$, $T=1\:\textrm{K}$, $a\tau_{0}\approx4.7\times10^{3}$.
The effective mean free path versus dimensionless frequency $x$.
For each of the fixed value of the turbulence correlation length $l_{turb}/\lambda_{mfp}=10^{-2},10^{-1},1,10,10^{2},10^{3},10^{4}$
the corresponding subgraph shows the effective mean three path as
a function of $x$ for the following values of the turbulent velocity:
$v_{turb}/v_{th}=0.25,0.5,1,2.5,5,10,10^{2}$, as well as for the
case with no turbulence. $T=1\:\textrm{K}$. The parameters are $\tau_{0}=10^{5}$,
$T=1\:\textrm{K}$, $a\tau_{0}\approx4.7\times10^{3}$.}
\end{figure*}

\newpage{}
\begin{figure*}[H]
\includegraphics[width=0.3\paperwidth]{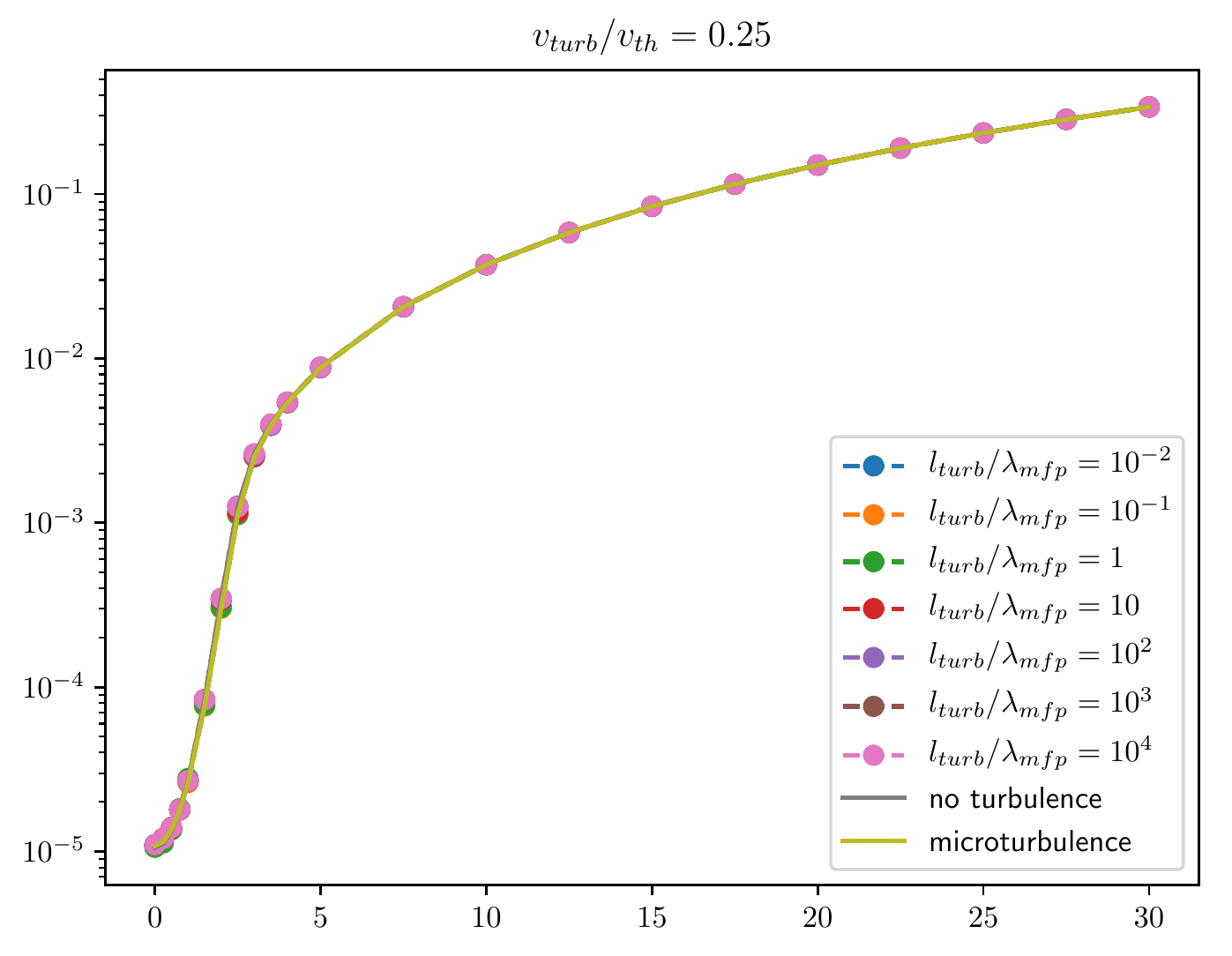}\includegraphics[width=0.3\paperwidth]{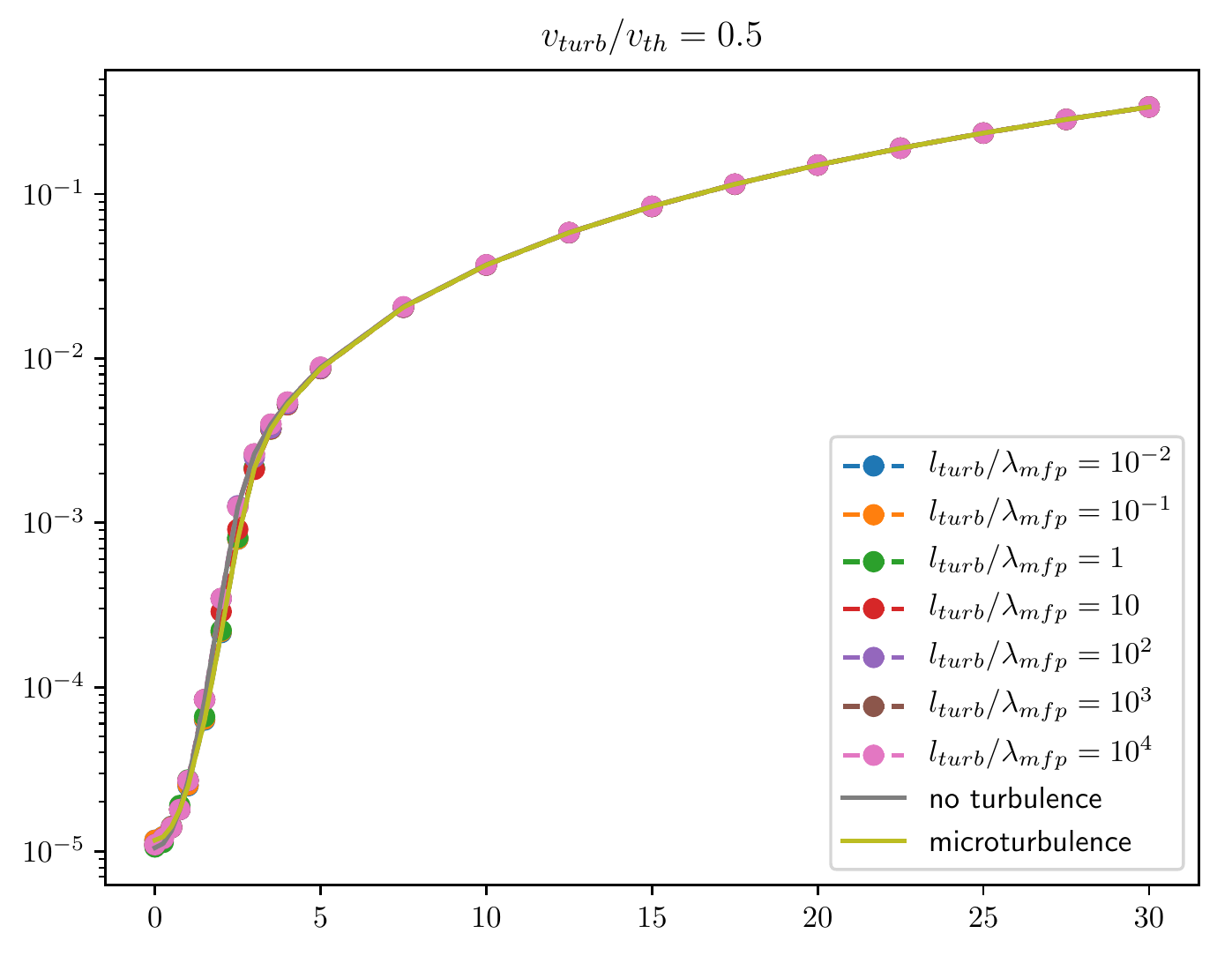}\includegraphics[width=0.3\paperwidth]{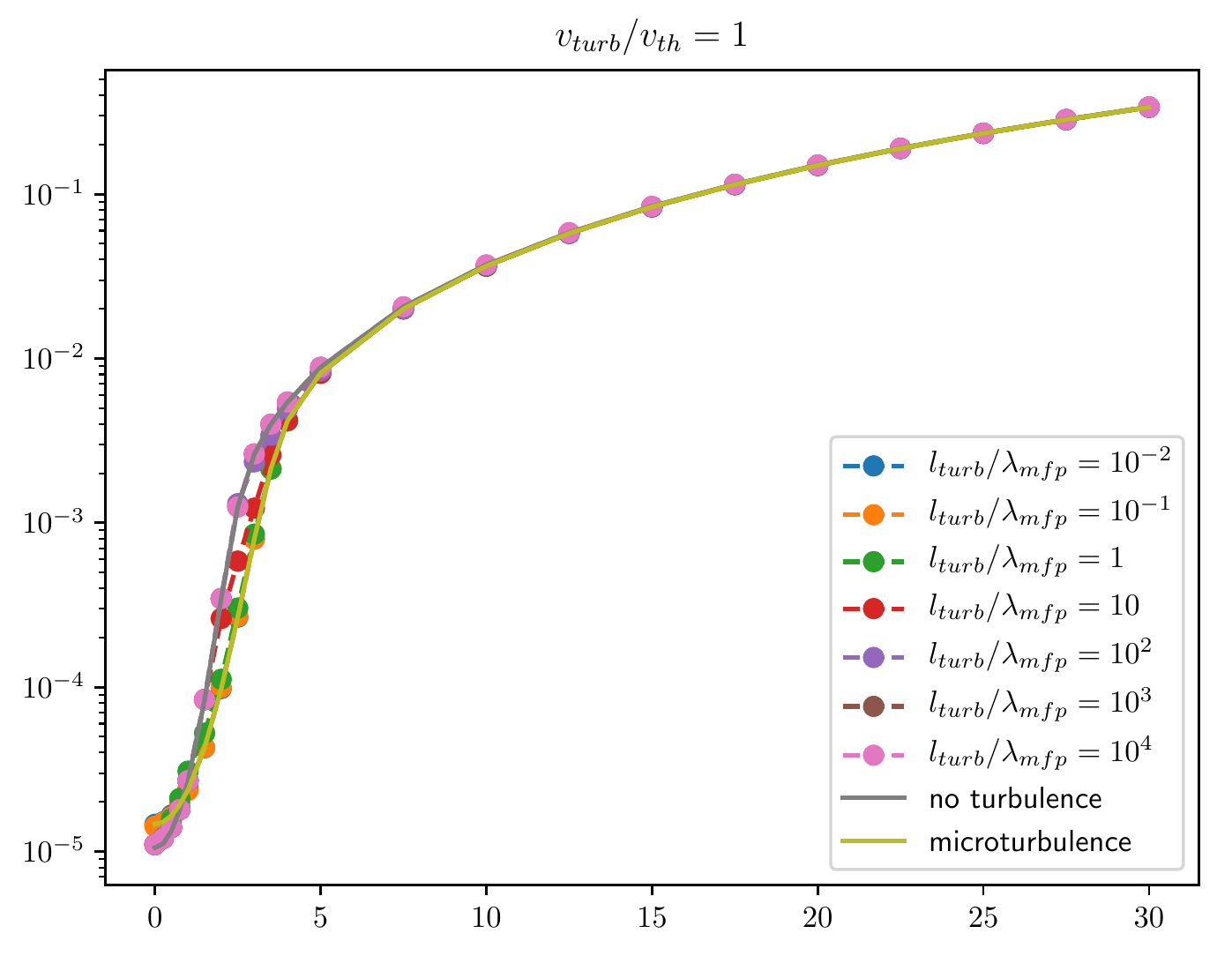}
\includegraphics[width=0.3\paperwidth]{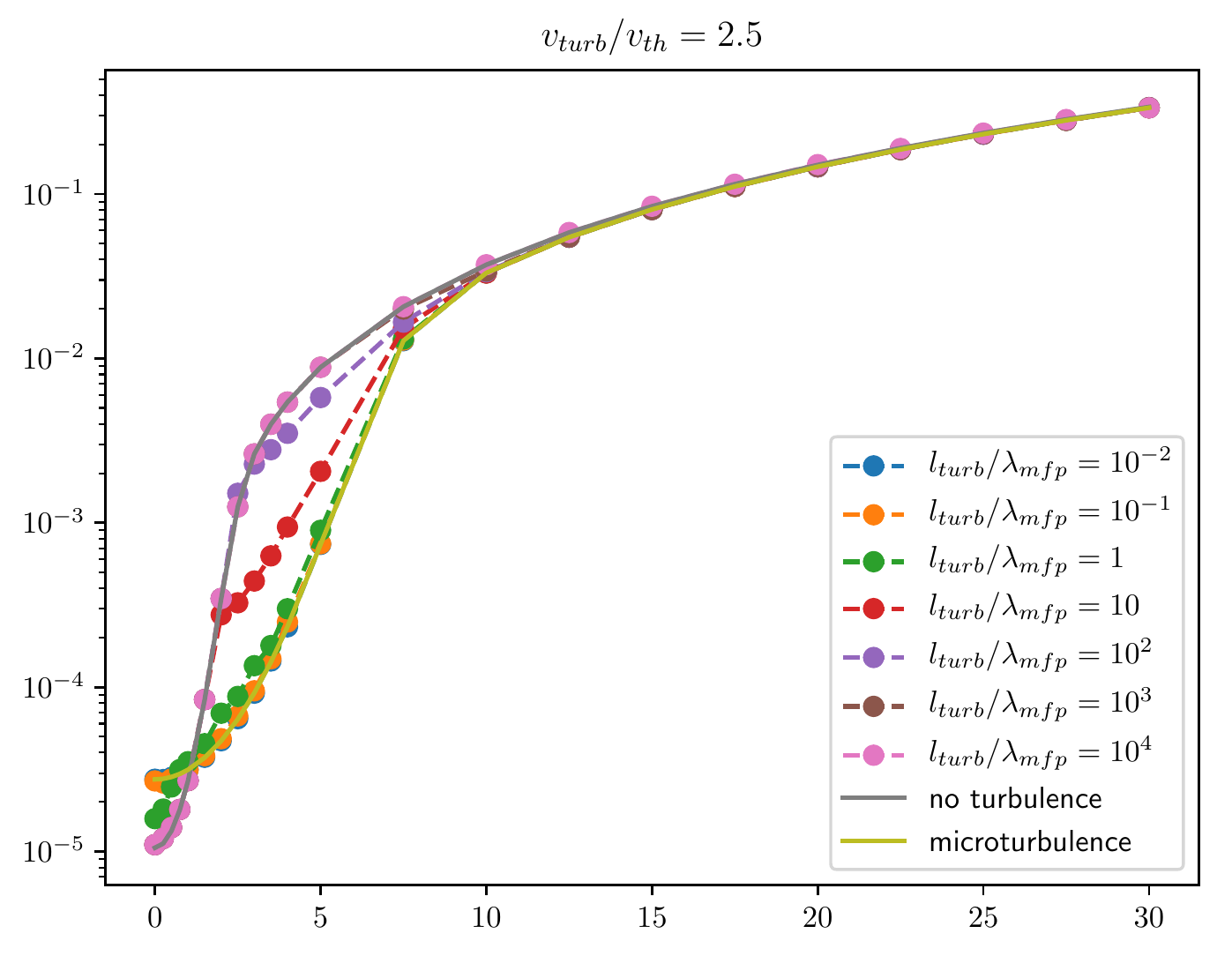}\includegraphics[width=0.3\paperwidth]{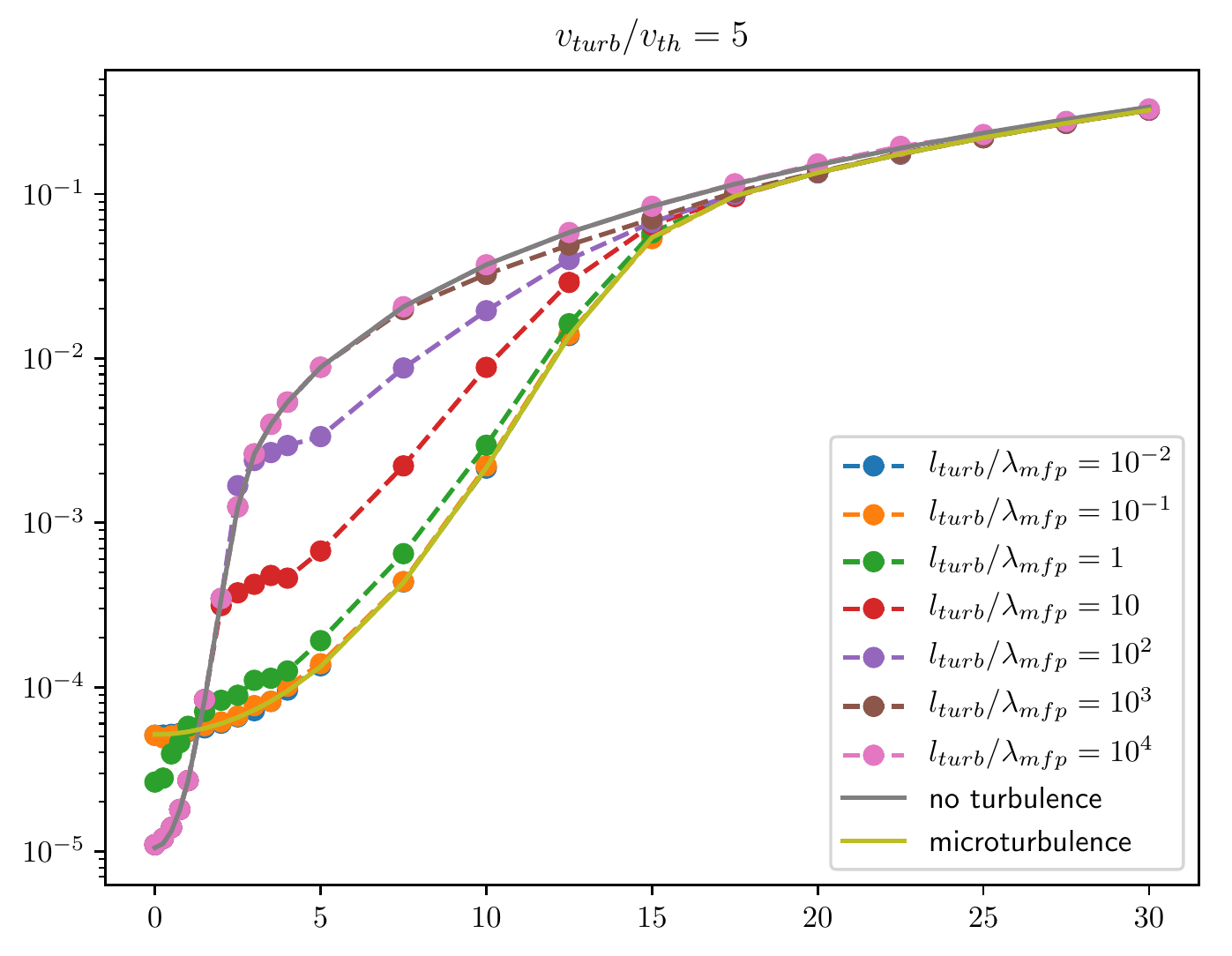}\includegraphics[width=0.3\paperwidth]{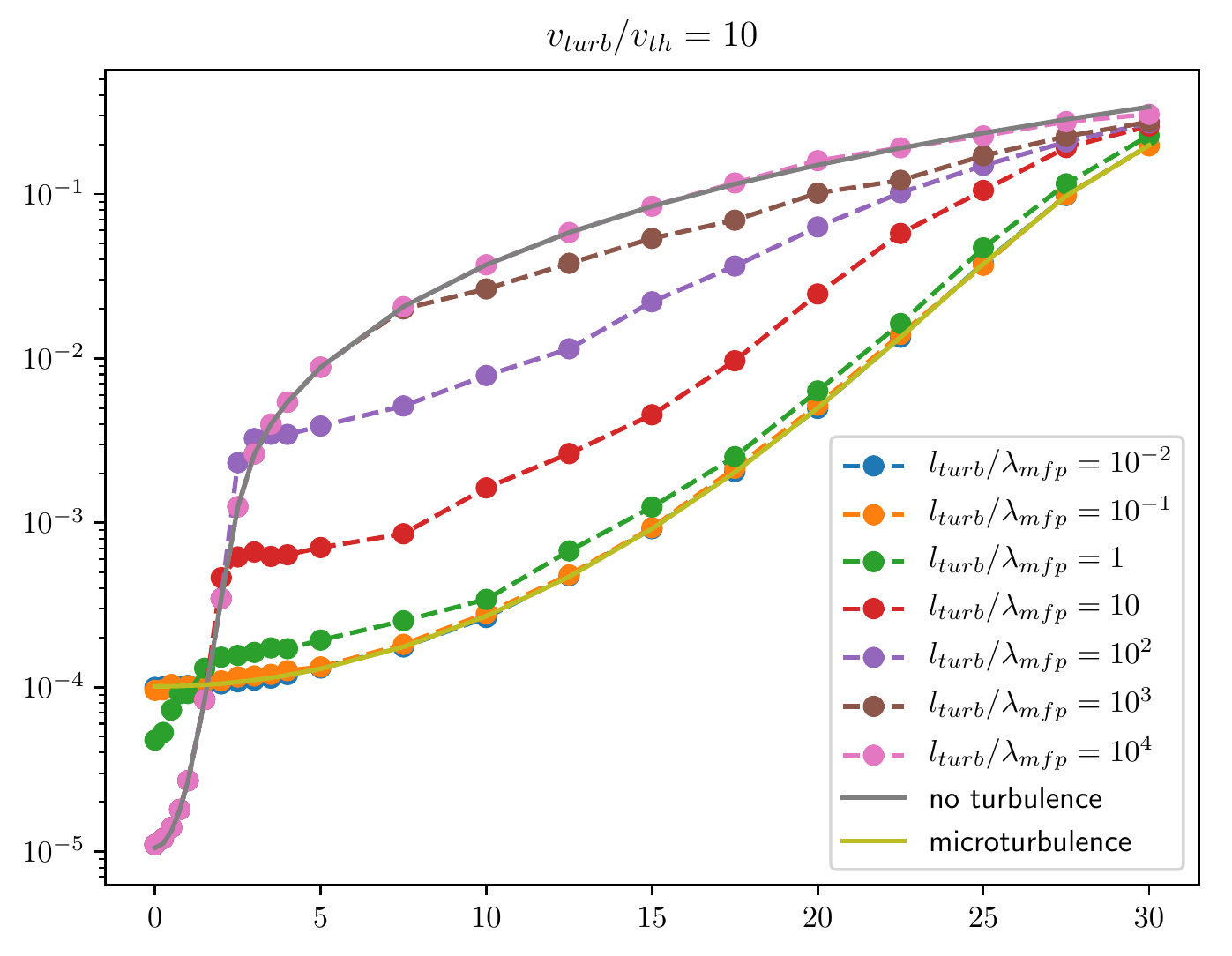}
\includegraphics[width=0.3\paperwidth]{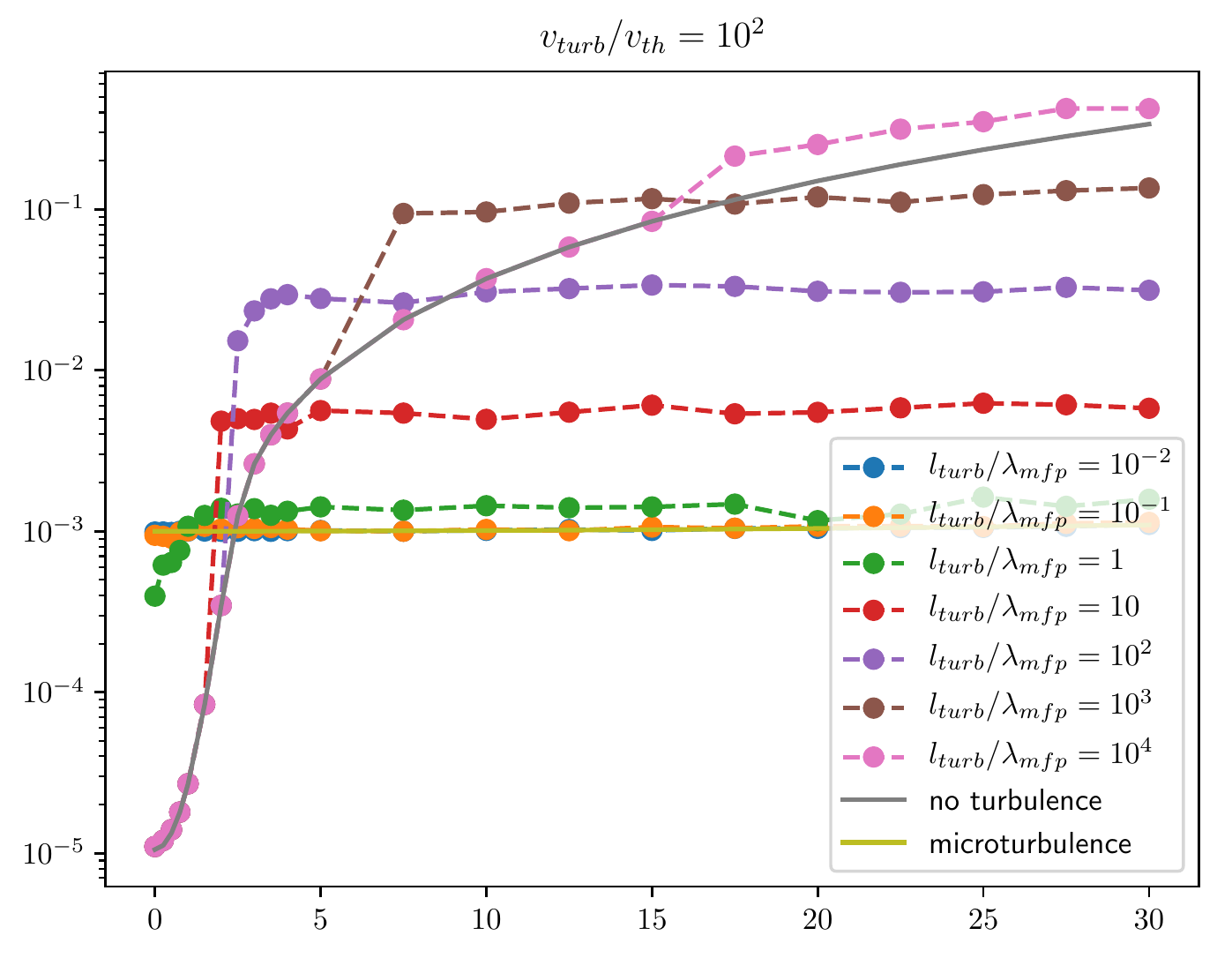}\includegraphics[width=0.3\paperwidth]{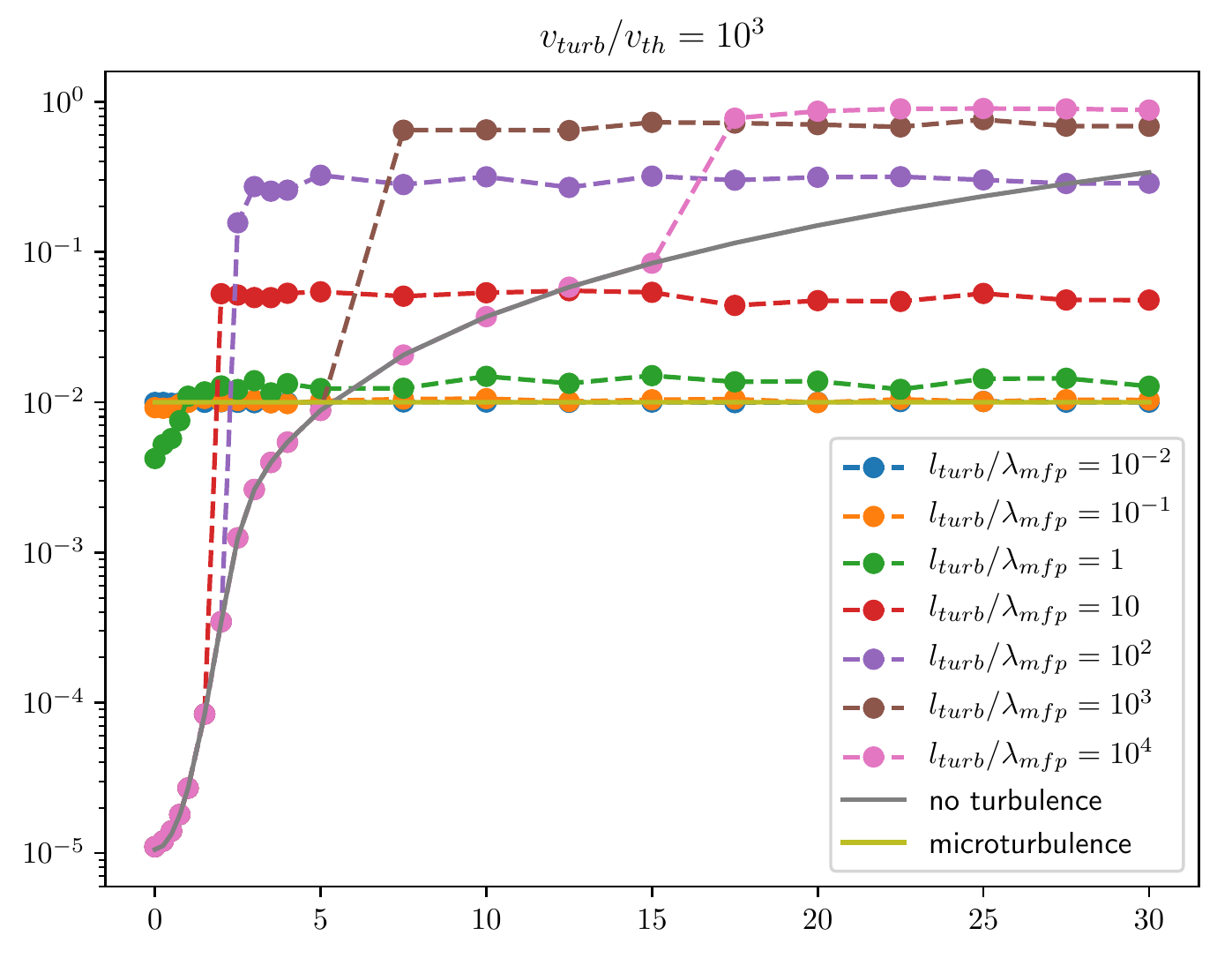}
\caption{The numerically calculated effective mean free path versus the dimensionless
frequency $x$ for $\tau_{0}=10^{5}$, $T=1\:\textrm{K}$, $a\tau_{0}\approx4.7\times10^{3}$.
The effective mean free path versus dimensionless frequency $x$.
For each of the fixed value of the turbulent velocity ($v_{turb}/v_{th}=0.25,0.5,1,2.5,5,10,10^{2}$)
the corresponding subgraph shows the effective mean three path as
a function of $x$ for the following values of the turbulence correlation
length ($l_{turb}/\lambda_{mfp}=10^{-1},1,10,10^{2},10^{3},10^{4}$)
as well as for the cases with no turbulence and with microturbulence.
The parameters are $\tau_{0}=10^{5}$, $T=1\:\textrm{K}$, $a\tau_{0}\approx4.7\times10^{3}$.}
\end{figure*}

% Don't change these lines
\bsp	% typesetting comment
\label{lastpage}
\end{document}